\documentclass[a4paper,11pt]{article}
\usepackage{jheppub} 
\usepackage{lineno}
\usepackage{placeins} 
\usepackage{subcaption}
\usepackage{capt-of}  
\usepackage[percent]{overpic}
\usepackage{xcolor}
\usepackage{graphicx}
\usepackage{subcaption}

\newcommand{\eq}{\begin{equation}}
\newcommand{\eqe}{\end{equation}}
\newcommand{\eqa}{\begin{eqnarray}}
\newcommand{\eqae}{\end{eqnarray}}

\title{\boldmath Sampling the Graviton Pole and Deprojecting the Swampland
}

\author{Guangzhuo Peng$^{a,b,c}$,
Laurentiu Rodina$^{d}$,
Anna Tokareva$^{a,f,e}$,
Yongjun Xu$^{a,b,c}$
}

\affiliation{
$^{a}$ School of Fundamental Physics and Mathematical Sciences,\\
Hangzhou Institute for Advanced Study, UCAS, Hangzhou 310024, China
\\[1mm]
$^{b}$ Institute of Theoretical Physics, Chinese Academy of Sciences,
Beijing 100190, China
\\[1mm]
$^{c}$ University of Chinese Academy of Sciences,
Beijing 100049, China
\\[1mm]
$^{d}$ Beijing Institute of Mathematical Sciences and Applications (BIMSA),
Beijing 101408, China
\\[1mm]
$^{f}$ International Centre for Theoretical Physics Asia-Pacific,
Beijing / Hangzhou, China
\\[1mm]
$^{e}$ Department of Physics, Blackett Laboratory, Imperial College London,\\
London SW7 2AZ, United Kingdom
}

\emailAdd{pengguangzhuo23@mails.ucas.ac.cn,  laurentiu.rodina@gmail.com, tokareva@ucas.ac.cn, xuyongjun23@mails.ucas.ac.cn}

\abstract{
We develop a primal bootstrap framework for effective field theories in the presence of a graviton pole, based on finite-resolution sampling rather than smearing, while also allowing direct control over the number of subtractions. We show that this approach reproduces the known projective bounds obtained from smearing in $D{\ge}6$, while yielding slightly stronger bounds in $D{=}5$. This method also makes it straightforward to impose linearized unitarity directly and provides an access to the extremal spectra. Applying the method to crossing-symmetric dispersion relations, we derive new non-projective bounds that fix the overall scale of the EFT couplings. In $D{=}5$, for example, we find that $\frac{M}{M_{\rm P}}{\lesssim}7.8$, showing that the EFT cutoff cannot be taken parametrically larger than the Planck scale. At the extremal values of the couplings, the spectra exhibit a surprising structure: for projective bounds, they exhibit peaks around quadratic Regge-like trajectories, while for the non-projective bounds they form sharp quadratic bands. In the latter case, the leading coefficients further display an inverse-quadratic dependence on the band number.}

\begin{document}
\maketitle
\flushbottom
\section{Introduction}
Effective field theory (EFT) provides the natural language for describing
fundamental physics at accessible energies. In this framework, physical
observables are organized as an expansion in energies and momenta divided
by a characteristic ultraviolet scale $M$. Locality, relativistic
causality, unitarity, and symmetry principles ensure that this expansion
is predictive and systematically improvable. Locality implies that the
dynamics can be encoded in a local action built from a finite set of
degrees of freedom, while dimensional analysis organizes the EFT into a
hierarchy of higher-derivative operators. Each operator is accompanied by
an a priori independent Wilson coefficient, encoding information about
ultraviolet physics.

A central question is therefore: \emph{which values of Wilson coefficients
are compatible with fundamental principles?} Dispersive arguments provide
powerful and model-independent answers. Under mild assumptions of
analyticity, unitarity, and locality, forward dispersion relations imply
positivity constraints on broad classes of Wilson coefficients
\cite{Adams:2006sv}. Incorporating crossing symmetry further strengthens
these results, leading to two-sided bounds consistent with dimensional
analysis and largely independent of the ultraviolet completion
\cite{Tolley:2020gtv,Caron-Huot:2020cmc}. This program has by now yielded a broad range of constraints on consistent EFTs across many settings---from chiral Lagrangians and scalar EFTs to gauge theories, gravitational EFTs, SMEFT, and scalar-tensor systems---using a correspondingly diverse set of tools, including dispersive positivity bounds, crossing-symmetric sum rules, dual and primal bootstrap methods, positive geometries or moment-problem formulations~\cite{Pham:1985cr, Pennington:1994kc,Nicolis:2009qm, Komargodski:2011vj,Remmen:2019cyz,Bellazzini:2019xts,Herrero-Valea:2019hde,Bellazzini:2020cot,Bellazzini:2017fep,Alberte:2020jsk,deRham:2017avq,deRham:2017zjm,deRham:2017imi,Wang:2020jxr,Alberte:2020bdz, Tokuda:2020mlf,Li:2021lpe,Caron-Huot:2021rmr,Du:2021byy,Bern:2021ppb,Li:2022rag, Caron-Huot:2022ugt,Saraswat:2016eaz,Arkani-Hamed:2021ajd,Herrero-Valea:2020wxz,Guerrieri:2021ivu,Henriksson:2021ymi,EliasMiro:2022xaa,Bellazzini:2021oaj, Herrero-Valea:2022lfd,Hong:2023zgm,Chiang:2022jep,Huang:2020nqy,Noumi:2021uuv, Xu:2023lpq, Chen:2023bhu,Noumi:2022wwf,deRham:2022hpx, Hong:2024fbl,Bern:2022yes,Ma:2023vgc,DeAngelis:2023bmd,Acanfora:2023axz,Aoki:2023khq,Xu:2024iao,EliasMiro:2023fqi,McPeak:2023wmq,Riembau:2022yse,Caron-Huot:2024tsk,Caron-Huot:2024lbf,Wan:2024eto,Berman:2024owc,Beadle:2024hqg,deRham:2025vaq,Bellazzini:2025shd,Ye:2025zhs,Bonnefoy:2025uzf,Bellazzini:2025bay,Gumus:2025hwq,Fernandez:2026fby}.

Once dynamical gravity is included, these questions take on a qualitatively different character. Broad swampland considerations suggest that not every
apparently consistent low-energy EFT can arise from quantum gravity
\cite{Vafa:2005ui,Ooguri:2006in,Brennan:2017rbf,Palti:2019pca,vanBeest:2021lhn}.
More concretely, ideas such as the Weak Gravity Conjecture
\cite{Arkani-Hamed:2006emk} and the Completeness Hypothesis
\cite{Banks:2010zn,Banks:1988yz,Polchinski:2003bq} point toward the
expectation that gravity places sharp and intrinsically nontrivial
constraints on the spectrum and interactions of any effective theory.
From this perspective, the S-matrix bootstrap provides a particularly
natural framework: it allows one to formulate these questions directly in
terms of general consistency conditions---analyticity, unitarity,
crossing symmetry, and causality---without committing to a specific
microscopic UV completion. In this sense, bounding gravitational EFT data
from the S-matrix is not merely a technical exercise, but a concrete and
model-independent route toward characterizing the boundary between the
Landscape and the Swampland.

However, with gravity, the standard dispersive analysis encounters a new technical obstruction. Forward scattering amplitudes are dominated by $t$-channel graviton exchange, leading to infrared singularities that invalidate the standard positivity arguments based on fixed-$t$ dispersion relations. In Ref.~\cite{Caron-Huot:2021rmr} this problem was overcome by reformulating the dispersion relations in terms of smearing over wavepackets localized in impact-parameter space.

While conceptually powerful, the smearing approach is not well suited to
implementing linearized unitarity beyond positivity alone. Linearized unitarity
is a pointwise constraint in partial-wave space,
\[
0 \le \rho_J(s) \le 2,
\]
whereas smearing introduces nonlocal correlations that obscure this linear
structure and complicate the numerical implementation. This limitation is
closely tied to the geometry of EFT coupling space. With positivity alone, the
constraints are homogeneous in the spectral density and define a projective
convex cone---the EFT-hedron---so that only ratios of Wilson coefficients are
bounded \cite{Arkani-Hamed:2020blm,Chiang:2021ziz}. Once the upper bound
$\rho_J(s)\le 2$ is imposed, this scaling is broken and the cone is deprojected
into a bounded, non-projective region \cite{Chiang:2022ltp,Chiang:2022jep}. In
the presence of gravity this deprojection is especially significant, because the
gravitational coupling itself is a physical parameter that cannot be scaled
away.

Motivated by these observations, we develop an alternative \emph{primal} bootstrap framework that can overcome the gravitational pole, based on sampling the functional constraints at finite kinematic resolution. This perspective allows us to impose positivity and linearized unitarity in a direct and numerically tractable way, and to derive absolute, non-projective bounds in gravitational EFTs. A further intrinsic advantage of the primal approach is that it automatically returns the extremal spectra. Our implementation differs from the "$\rho$ primal bootstrap" of \cite{Guerrieri:2021ivu}, which effectively assumes improved UV behavior and, correspondingly, a zero-subtraction dispersion relation. By contrast, in our approach the number of subtractions is an explicit input and can be controlled exactly.\\

Concretely, we start from the fixed-$t$, $k=2$ improved sum rules of Refs.~\cite{Caron-Huot:2021rmr}, which relate the low-energy EFT data to an integral over the UV spectral density. Schematically, they take the form
\eq
\mathrm{IR}_{k=2}(t)=\mathrm{UV}_{k=2}(t),
\eqe
namely
\begin{equation}
\begin{aligned}\label{disp}
\frac{8\pi G}{-t} + 2 g_2 - g_3 t
&=
\left\langle
\frac{(2s' + t)\, P_J\!\left(1 + \frac{2t}{s'}\right)}
     {s' (s' + t)^2}
-
\frac{t^2}{(s')^3}
\left(
\frac{(4s' + 3t)\, P_J(1)}{(s' + t)^2}
+
\frac{4t\, P_J'(1)}{(s')^2 - t^2}
\right)
\right\rangle
\\
&\equiv
\left\langle C^{\mathrm{improved}}_{2,t}[s',t,J] \right\rangle ,
\end{aligned}
\end{equation}
where
\begin{equation}
\bigl\langle \cdots \bigr\rangle
\equiv
\sum_{J=0}^{\infty}
n_J^{(D)}
\int_{M^2}^{\infty}
\frac{\mathrm ds'}{\pi}\,
{s'}^{\frac{2-D}{2}}\,
\rho_J(s')\,
(\cdots).
\end{equation}
Our strategy is then simply to impose this dispersive relation at a finite set of kinematic points $t_i\in(-1,0)$,
\eq
\mathrm{IR}_2(t_i)=\mathrm{UV}_2(t_i)\,, \qquad i=1,\dots,N_t,
\eqe
thereby turning the continuous bootstrap constraint into a finite system suitable for the primal optimization problem.

These are manifestly \emph{necessary} conditions. In the limit of sufficiently dense sampling, one expects this procedure to converge, producing robust and optimal bounds that are equivalent to smearing.

Given the resulting set of $N_t$ equations, we solve the constraints using a primal
bootstrap approach \cite{Chiang:2022jep}. The basic idea is to discretize the dispersive energy integral into $N_\mu$ points, truncate the
spin sum to $J_\textrm{max}$, and treat the resulting finite collection of spectral densities as optimization
variables. Positivity or linearized unitarity can then be imposed directly as linear
constraints, and higher-$k$ sum rules or null constraints can be incorporated simply by
adding further linear relations. In this way the problem reduces to a finite linear program,
whose convergence can be tested by systematically increasing the energy resolution, the
number of sampling points, and the spin cutoff. As an added advantage, the optimization program directly yields the explicit extremal spectra.

While the sampling approach is conceptually simple, its numerical implementation is delicate, which likely explains why it has not been pursued previously in the presence of a graviton pole. The difficulty is twofold: a primal linear-programming formulation is generally less stable than dual methods, and the graviton pole forces one to sample near a singular kinematic region. As a result, a naive implementation leads to poor convergence and severe numerical instabilities.

We show that these problems can nevertheless be controlled. The essential ingredients are to use Chebyshev nodes that sample more densely towards the singular endpoints, and to correlate the number of sampling points with the spin cutoff $J_{\max}$ so as to remain in a stable truncation window. With these choices, the method becomes robust. We find, however, that the fixed-$t$ formulation still converges relatively slowly and is therefore not well suited to extracting non-projective bounds. For this reason, we turn to the manifestly crossing-symmetric fixed-$a$ dispersion relations ~\cite{Mahoux:1974ej,Auberson:1972prg,Sinha:2020win}, which are numerically more stable, converge faster, and yield both non-projective bounds and the corresponding extremal spectra.

\subsection{Main results}

\begin{figure}[t]
\centering
\includegraphics[width=1\textwidth]{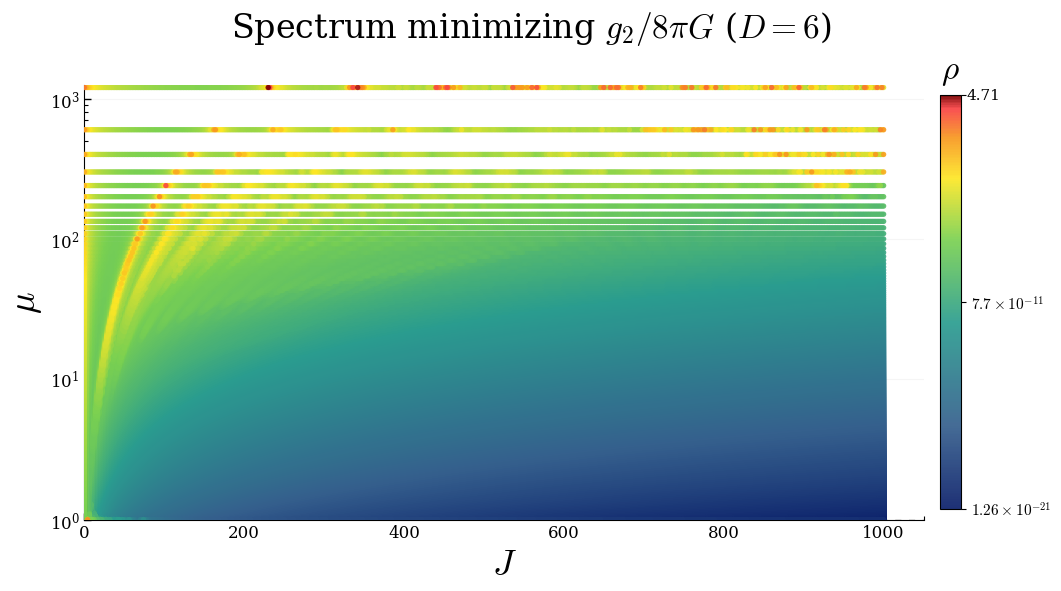}
\caption{Log plot of
extremal spectral density $\rho_J(\mu)$ in $D=6$ at the projective lower bound of $g_2/8\pi G$. The extremal spectrum shows quadratic Regge-like trajectories in the $(J,\mu)$ plane. Away from these trajectories, the spectral density is strongly suppressed. States with both large mass and spin also tend to carry enhanced spectral weight. The result is obtained with $N_t=60$, $J_{\max}=1000$, and $N_\mu=1200$.}
\label{fig:extremal_spectrum_6d_intro1}
\end{figure}

\begin{figure}[t]
\centering
\includegraphics[width=1\textwidth]{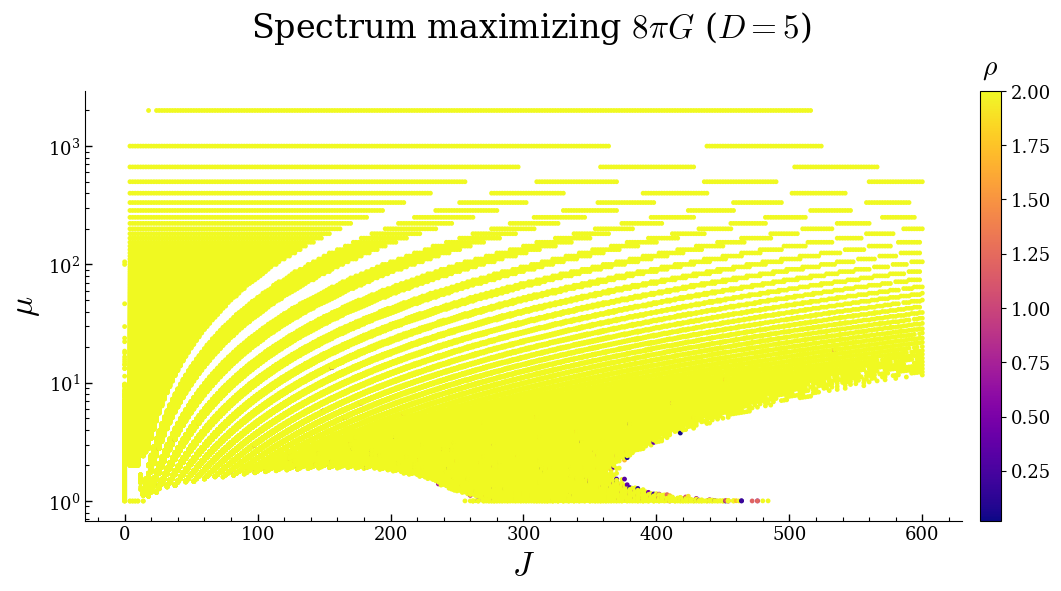}
\caption{Log plot for the 
extremal spectral density $\rho_J(\mu)$ in $D=5$ at the non-projective upper bound on $8\pi G$. The extremal solution has a sharp structure ($\rho$ is mostly either exactly 0 or 2) and organizes into multiple
quadratic Regge-like bands in the $(J,\mu)$ plane. The result is obtained with $N_a^{(2)}=200$, $J_{\max}=600$, $N_\mu=2000$.}
\label{fig:extremal_spectrum_5d_non}
\end{figure}

Our main findings are as follows. 

\paragraph{Fixed-$t$ and projective bounds}
First, we find fixed-$t$ sampling reproduces the known projective bounds in $D\ge 6$, providing a nontrivial validation of the method. In $D=5$ we find a deviation from the smeared result, implying a slightly stronger bound of $g_2/8\pi G\ge -16.5$  compared to $g_2/8\pi G\ge -18$ from smearing.

Our method also allows a straightforward extraction of the extremal spectrum. In Fig.~\ref{fig:extremal_spectrum_6d_intro1} we present the extremal spectral density at the lower bound of $g_2/8\pi G$ in $D=6$. Although the bound is close to that obtained in Ref.~\cite{Caron-Huot:2021rmr} ($-9.77$ vs.\ $-9.57$), the extremal spectrum is qualitatively different. In the primal bootstrap, we find states populate the full allowed region but most carry parametrically suppressed spectral weight $\rho$. The dominant support instead organizes into quadratic Regge-like trajectories in the $(J,\mu)$ plane, with strong suppression away from them. This contrasts with the dual bootstrap result in Ref.~\cite{Caron-Huot:2021rmr}, where the spectrum was confined to a lower-right triangular region with isolated states.

\paragraph{Fixed-$a$ and non-projective bounds}
Second, we find that fixed-$a$ sampling is numerically more stable and yields projective bounds close to those obtained from smeared fixed-$a$ dispersion relations. Once linearized unitarity is imposed, fixed-$a$ leads to genuinely new
non-projective bounds in gravitational EFTs for $D\ge 5$. In particular, in $D=5$ we
obtain an upper bound on the dimensionless gravitational coupling, equivalent to
\eq\label{eq:EFT_breakdown_5d}
0<\frac{M}{M_{\rm P}}\lesssim 7.8\, .
\eqe
We plot the extremal spectrum corresponding to this bound in Fig.~\ref{fig:extremal_spectrum_5d_non}. We observe the spectrum organizes into sharp bands in the $(J,\mu)$ plane, whose boundaries are well fit by quadratic trajectories
\begin{equation}
\mu_n\sim b_{0,n}+b_{1,n} J+ b_{2,n}J^2\,,
\end{equation}
with the leading coefficients themselves following an approximate inverse-quadratic
pattern in the band number,
\eq
b_{2,n} \sim \frac{A}{(B+n)^2}.
\eqe
Similar higher-dimensional spectra are shown in Fig.~\ref{fig:spectrum_d67}.

A striking feature of our extremal spectra is the emergence of quadratic Regge-like trajectories in the $(J,\mu)$ plane. This is unexpected in light of previous studies of the S-matrix bootstrap, where linear Regge trajectories have been the dominant structure. Such linear behavior has been extensively explored, either as part of the input data---for instance through assumptions about the exchanged spectrum or string-inspired Regge behavior---or as a property emerging in extremal solutions with string-like characteristics
\cite{Nayak:2017qru,Haring:2023zwu,Eckner:2024ggx,Bose:2020cod,Albert:2023seb,Albert:2024yap,Eckner:2024pqt,Cheung:2024uhn,Cheung:2024obl,Berman:2024wyt,Bhat:2024agd,Huang:2025icl}. It is also worth noting that Ref.~\cite{Caron-Huot:2021rmr} observed approximately linear trajectories in spectra saturating projective bounds. 

By contrast, our extremal spectra exhibit a quadratic organization rather than linear trajectories. Notably, this structure arises dynamically, without imposing any additional assumptions on the ultraviolet behavior of the theory.

\paragraph{Organization of the paper.}
The paper is organized as follows. In
Section~\ref{sec:unitarity_crossing}, we review partial-wave unitarity and
crossing-symmetric dispersion relations. In Section~\ref{sec:distributions}, we introduce the approach based on sampling in either $t$ or $a$, as well as the primal bootstrap approach to solve the constraints.  Section~\ref{sec:4d_nogravity} presents illustrative results in four dimensions without gravity. Section~\ref{sec:5d6d_gravity} contains our main results in the presence of gravity, including projective and non-projective bounds, and extremal spectra. Appendix~\ref{app:numerical_stability} collects numerical details and
stability tests. We discuss loop corrections in Appendix~\ref{loopchecks}, and in Appendix~\ref{sec:4d_gravity} we discuss the logarithmic behavior of bounds in $D=4$ gravity as reflected in the sampling approach.

\section{Review of dispersion relations}
\label{sec:unitarity_crossing}
\subsection{Partial-wave expansion and unitarity}
\label{subsec:partialwave}

We consider the $2\to2$ scattering amplitude $\mathcal M(s,t)$ for identical,
massless scalar particles in $D$ spacetime dimensions.
Assuming Lorentz invariance and crossing symmetry, the amplitude admits
the partial-wave expansion
\begin{equation}
\mathcal M(s,t)
=
\sum_{J=0}^{\infty}
n_J^{(D)}\,
a_J(s)\,
P_J^{(D)}\!\left(1+\frac{2t}{s}\right),
\label{eq:partialwave}
\end{equation}
where $P_J^{(D)}(x)$ are the Gegenbauer polynomials appropriate to $D$
dimensions, and the normalization factor is given by
\begin{equation}
n_J^{(D)}
=
\frac{(4\pi)^{\frac{D}{2}}\,(D+2J-3)\,\Gamma(D+J-3)}
{\pi\,\Gamma\!\left(\frac{D-2}{2}\right)\,\Gamma(J+1)} .
\label{eq:normfactor}
\end{equation}
With this choice of normalization, the unitarity constraints take a
standard form in terms of the partial-wave coefficients $a_J(s)$.

Analyticity of the $S$-matrix implies that each $a_J(s)$ admits a
spectral representation with a positive spectral density.
It is convenient to define the dimensionless quantity
\begin{equation}
\rho_J(s)
\equiv
s^{\frac{D-4}{2}}\,\mathrm{Im}\,a_J(s),
\label{eq:spectral_density}
\end{equation}
where the prefactor removes the $D$-dimensional phase-space scaling.
Unitarity then implies the bounds
\begin{equation}
0 \le \rho_J(s) \le 2 .
\label{eq:rho_bounds}
\end{equation}
The lower bound follows from positivity of the spectral density, while the
upper bound arises from linearized unitarity, which is appropriate in
weakly coupled or perturbative regimes.

Assuming analyticity and polynomial boundedness (see \cite{Haring:2022cyf} for a recent discussion in the specific case of gravity), the scattering amplitude
satisfies a twice-subtracted dispersion relation in the $s$-channel,
\begin{equation}
\mathcal M(s,t)
=
\mathcal M(0,t)
+
s\,\left.\frac{\partial \mathcal M(s,t)}{\partial s}\right|_{s=0}
+
\frac{s^2}{\pi}
\int_{M^2}^{\infty}
\frac{\mathrm ds'}{s'^2}\,
\frac{\mathrm{Im}\,\mathcal M(s',t)}{s'-s} ,
\label{eq:fixedt_DR}
\end{equation}
where $M^2$ denotes the lightest physical threshold.
The twice-subtracted form is required by locality, which implies the
high-energy behavior
\begin{equation}
\lim_{t\to0^-}\frac{\mathcal M(s,t)}{s^2}=0 .
\label{eq:locality_condition}
\end{equation}

The $D$-dimensional Gegenbauer polynomials appearing in
\eqref{eq:partialwave} admit the hypergeometric representation
\begin{equation}
P_J^{(D)}(x)
=
{}_2F_1\!\left(
-J,\,
J+D-3;\,
\frac{D-2}{2};\,
\frac{1-x}{2}
\right).
\label{eq:gegenbauer}
\end{equation}

Finally, it is useful to introduce the positive spectral average
\begin{equation}
\bigl\langle \cdots \bigr\rangle
\equiv
\sum_{J=0}^{\infty}
n_J^{(D)}
\int_{M^2}^{\infty}
\frac{\mathrm ds'}{\pi}\,
{s'}^{\frac{2-D}{2}}\,
\rho_J(s')\,
(\cdots),
\label{eq:positiveaverage}
\end{equation}
which is manifestly non-negative due to $2\geq\rho_J(s')\ge0$.
All dispersive constraints derived below can be expressed in terms of
such positive averages.

\subsection{Fixed-$t$ dispersion relations and improved sum rule}
Under standard assumptions of analyticity, unitarity, crossing symmetry,
and bounded high-energy growth, the scattering amplitude $\mathcal M(s,t)$
satisfies a twice-subtracted fixed-$t$ dispersion relation for $t<0$,
\begin{equation}
\frac{\mathcal M(s,t)}{s(s+t)}
=
\int_0^\infty \frac{ds'}{\pi}
\left(
\frac{1}{s'-s} + \frac{1}{s'+s+t}
\right)
\operatorname{Im}
\!\left[
\frac{\mathcal M(s',t)}{s'(s'+t)}
\right].
\label{eq:fixedt_dispersion}
\end{equation}
This representation follows from a contour deformation in the complex
$s$-plane and makes crossing symmetry between the $s$- and $u$-channels
manifest.

\paragraph{Arc integral at fixed $t$}
\begin{figure}[t]
\centering
\includegraphics[width=0.6\textwidth]{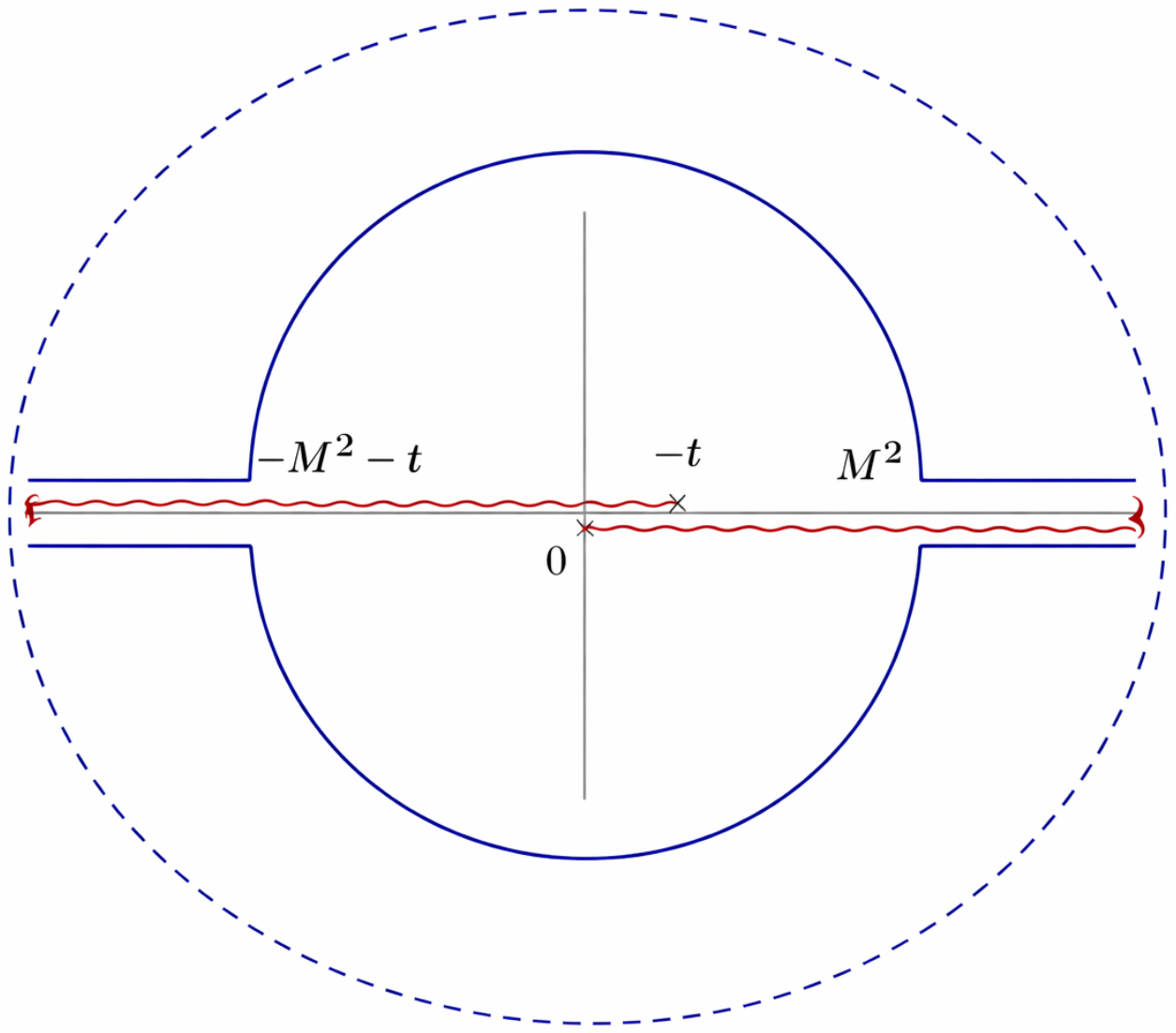}
\caption{
Contour used in the fixed-$t$ dispersion relation.
The integration consists of a low-energy arc enclosing the origin
and integrals along the branch cuts on the real axis.
The contribution from the large circular arc for $k\geq2$ at $|s|\to\infty$
vanishes as a consequence of locality.
}
\label{fig:fixt}
\end{figure}

To extract low-energy constraints, the dispersion integral is split at a
scale $M^2$ into low- and high-energy contributions. The low-energy part
can be rewritten as an integral over a finite arc in the complex plane and
evaluated using the EFT expansion, while the high-energy contribution
admits a partial-wave representation with positive spectral weights. Their
difference defines a dispersive quantity $B(s,t)$, which is analytic around
$s=0$ and $s=-t$.

Expanding $B(s,t)$ at small $s$ and $s+t$ yields
\begin{equation}
B(s,t)
=
\sum_{k\,\mathrm{even}}
B_k(t)\,[s(s+t)]^{\frac{k}{2}-1},
\label{eq:Bexpansion}
\end{equation}
leading to an infinite family of sum rules labeled by the Regge spin $k$,
\begin{equation}
-\,B_k(t)\big|_{\mathrm{low}}
=
B_k(t)\big|_{\mathrm{high}},
\qquad k~\mathrm{even}.
\label{eq:basic_sumrule}
\end{equation}
The left-hand side involves Wilson coefficients of the EFT, while the
right-hand side is a positive average over partial waves weighted by known
kernels. These relations therefore impose nontrivial constraints on EFT
couplings.

A crucial observation is that crossing symmetry at low energies correlates
different sum rules \cite{Caron-Huot:2020cmc,Caron-Huot:2021rmr}. A single higher-derivative operator contributes
simultaneously to multiple $B_k(t)$. For example, the crossing-symmetric
interaction
\begin{equation}
\mathcal M \supset g_6 (s^2+t^2+u^2)^3
\end{equation}
feeds into the $k=2,4,6$ sum rules at the same order in the low-energy
expansion.

This redundancy enables the construction of special linear combinations of
sum rules—known as \emph{improved sum rules} or \emph{null constraints}—in
which the leading EFT contributions cancel identically at tree level.
Explicitly, improved sum rules are given by the following linear combinations of the
fixed-$t$ sum rules:
\begin{align}
B^{\mathrm{imp}}_2(t)
&=
B_2(t)
-2t^2 B_4(0)
+t^3 \partial_t B_4(0)
+3t^4 B_6(0)
+t^5 \partial_t B_6(0)
+\cdots
\nonumber\\[4pt]
&=
\oint \frac{ds}{4\pi i\, s}
\Bigg[
\frac{\mathcal M(s,t)}{s(s+t)}
-\frac{t^3 \partial_t \mathcal M(s,0)}{s^2 (s^2-t^2)}
+\frac{(2s^2-t^2)t^2\,\mathcal M(s,0)}{s^3 (s+t)(s^2-t^2)}
\Bigg],
\label{eq:B2imp}
\\[10pt]
B^{\mathrm{imp}}_4(t)
&=
B_4(t)
-2t^3 \partial_t B_6(0)
+\frac{1}{2}t^4 \partial_t^2 B_6(0)
+2t^4 B_8(0)
+\cdots
\nonumber\\[4pt]
&=
\oint \frac{ds}{4\pi i\, s}
\Bigg[
\frac{\mathcal M(s,t)}{s^2 (s+t)^2}
-\frac{t^4 \partial_t^2 \mathcal M(s,0)}{2 s^4 (s^2-t^2)}
+\frac{(2s^2-st-2t^2)t^3 \partial_t \mathcal M(s,0)}{s^5 (s+t)(s^2-t^2)}
\nonumber\\
&\hspace{4cm}
+\frac{(3s^3-4st^2-2t^3)t^2 \mathcal M(s,0)}{s^5 (s+t)^2 (s^2-t^2)}
\Bigg].
\label{eq:B4imp}
\end{align}

The tree-level EFT contributions to these improved sum rules indeed truncate
at low energies,
\begin{align}
-\,B^{\mathrm{imp}}_2(t)\big|_{\mathrm{low}}^{\mathrm{tree}}
&=
\frac{8\pi G}{-t}
+2 g_2
- g_3 t,
\label{eq:B2imp_tree}
\\[6pt]
-\,B^{\mathrm{imp}}_4(t)\big|_{\mathrm{low}}^{\mathrm{tree}}
&=
4 g_4
-2 g_5 t
+ g_6 t^2 ,\,
\label{eq:B4imp_tree}
\end{align}
leading to \eqref{disp}. Similarly, the lowest-spin null constraint can be written as
\begin{align}
X_4(t)
&=
B_4(t)
- B_4(0)
- t\,\partial_t B_4(0)
- \frac{1}{2} t^2 \partial_t^2 B_4(0)
- 2 t^3 \partial_t B_6(0)
- \frac{1}{2} t^4 \partial_t^2 B_6(0)
+\cdots
\nonumber\\[4pt]
&=
\oint \frac{ds}{4\pi i\, s}
\Bigg[
\frac{\mathcal M(s,t)}{s^2 (s+t)^2}
-\frac{t^2 \partial_t^2 \mathcal M(s,0)}{2 s^2 (s^2-t^2)}
-\frac{(s^2-st-t^2)t\,\partial_t \mathcal M(s,0)}{s^3 (s+t)(s^2-t^2)}
\nonumber\\
&\hspace{3.5cm}
-\frac{(s^2+st+t^2)(s^2-st-t^2)\mathcal M(s,0)}{s^4 (s+t)^2 (s^2-t^2)}
\Bigg],
\label{eq:X4_def}
\end{align}
and satisfies
\begin{equation}
X_4(t)\big|_{\mathrm{low}}^{\mathrm{tree}} = 0 .
\label{eq:X4_tree}
\end{equation}

Note that the null constraint $X_4(t)$ can equivalently be obtained from the
improved spin--$4$ sum rule via
\begin{equation}
X_4(t)
=
B^{\mathrm{imp}}_4(t)
- B^{\mathrm{imp}}_4(0)
- t\,\partial_t B^{\mathrm{imp}}_4(0)
- \frac{t^2}{2}\,\partial_t^2 B^{\mathrm{imp}}_4(0).
\end{equation}

While improved sum rules and null constraints provide a powerful way to eliminate
lower-dimensional EFT contributions and sharpen dispersive bounds, their construction
depends sensitively on the detailed low-energy polynomial expansion of the amplitude.
This makes them vulnerable to subleading EFT terms, to the order-by-order implementation
of crossing symmetry, and especially to loop effects. Loop corrections generically
introduce logarithms, scheme-dependent polynomial ambiguities, and potentially additional
infrared singularities, thereby spoiling the exact tree-level truncation properties on
which the improved sum rules rely. In practice, we also find that the fixed-$t$
formulation converges rather slowly and becomes numerically unstable as the ansatz
parameters are increased, particularly for non-projective bounds.

For these reasons, we will use fixed-$t$ mainly as a benchmark and then turn to
manifestly crossing-symmetric fixed-$a$ dispersion relations, which are numerically more
stable and better suited to extracting non-projective bounds. Loop effects in this
framework will be left for future work.
\subsection{Fixed-$a$ crossing-symmetric dispersion relations}
\label{subsec:crossing}

We now review the fully crossing-symmetric dispersion relations
developed in
Refs.~\cite{Mahoux:1974ej,Auberson:1972prg,Sinha:2020win},
restricting attention to amplitudes that are symmetric under permutations
of $(s,t,u)$.
For massless external states, $s+t+u=0$, and it is convenient to introduce
the crossing-invariant variables
\begin{equation}
x \equiv st+su+tu = -\frac12\left(s^2+t^2+u^2\right),
\qquad
y \equiv stu,
\qquad
a \equiv \frac{y}{x},
\label{eq:xa_def}
\end{equation}
so that the amplitude may be regarded as a function $\mathcal M(x,a)$.

As an illustration, the tree-level four-scalar scattering amplitude can be
written in the form
\begin{equation}
\mathcal M^{\mathrm{tree}}(x,a)
=
8\pi G\,\frac{x}{a}
+
\sum_{m,n} c_{mn}\,x^m a^n \, .
\label{eq:tree_xa}
\end{equation}
The pole at $a=0$ arises from long-range graviton exchange, while the
remaining terms correspond to local contact interactions in the effective
field theory. Locality imposes the constraint $n \le m$, which ensures that for any fixed
power of $x$ only a finite number of terms contribute to the expansion.

In terms of the variables $(x,a)$, the twice-subtracted,
crossing-symmetric dispersion relation can be written as
\begin{equation}
\mathcal M(x,a)
=
\alpha_0(a)
+
x
\int_0^\infty
\frac{\mathrm ds'}{\pi}\,
\frac{\mathrm{Im}\,\mathcal M\!\left(s',\tau(s',a)\right)}{(s')^3}\,
\frac{(2s'-3a)(s')^2}{x(a-s')-(s')^3},
\label{eq:crossing_DR}
\end{equation}
where $\alpha_0(a)$ is a subtraction function fixed by low-energy data.
The function
\begin{equation}
\tau(s',a)
=
-\frac{s'}{2}
\left(
1-\sqrt{\frac{s'+3a}{s'-a}}
\right)
\label{eq:tau_def}
\end{equation}
implements the inverse map from the crossing variable $a$ to the Mandelstam
variable $t$.
The two branches of the square root correspond to the two solutions related
by $t\leftrightarrow u$ crossing.

Expanding \eqref{eq:crossing_DR} around $x=0$ yields a family of dispersive
sum rules labeled by an even integer $k$,
\begin{equation}
-\,c_k^{\rm low}(a)
=
c_k^{\rm high}(a),
\qquad
k=2,4,6,\ldots,
\label{eq:sumrule}
\end{equation}
where only even powers appear due to crossing symmetry.
The high-energy contribution may be expressed in terms of partial waves as
\begin{equation}
c_k^{\rm high}(a)
=
\Biggl\langle
\frac{P_J^{(D)}\!\left(
\sqrt{\frac{s'+3a}{s'-a}}
\right)}{(s')^{3k/2}}\,
(s'-a)^{\frac{k}{2}-1}\,
(2s'-3a)
\Biggr\rangle .
\label{eq:ck_high}
\end{equation}
Since $2\ge\rho_J(s')\ge0$, the sign and magnitude of $c_k^{\rm high}(a)$ are
entirely controlled by the kernel appearing inside the brackets.

In the forward limit $|t|\ll s$, one has $a\simeq t$, and the
crossing-symmetric sum rules \eqref{eq:sumrule} reduce to the familiar improved
fixed-$t$ dispersion relations.
They may therefore be viewed as their fully crossing-symmetric
generalization.

\subsubsection*{Dispersive sum rules at fixed-$a$}
\label{subsec:fixeda_sumrules}

Starting from the crossing-symmetric dispersion relation
\eqref{eq:crossing_DR}, we obtain an infinite family of dispersive sum
rules by expanding around $x=0$ at fixed $a$.
To this end, we introduce the differential operator
\begin{equation}
\mathcal D_k
\equiv
\frac{(-1)^{k/2}}{(k/2)!}\,
\partial_x^{k/2},
\qquad
k=2,4,6,\ldots ,
\label{eq:Dk_def}
\end{equation}
and evaluate the dispersion relation at $x=0$.
This yields the sum rule
\begin{equation}
-\,c_k^{\rm low}(a)
=
c_k^{\rm high}(a),
\label{eq:fixeda_sumrule}
\end{equation}
where $k$ labels the order of the expansion and plays the role of an
effective Regge spin.

The high-energy contribution is given explicitly by
\begin{equation}
c_k^{\rm high}(a)
=
\int_{M^2}^{\infty}
\frac{\mathrm ds'}{\pi}\,
\mathrm{Im}\,\mathcal M\!\left(s',\tau(s',a)\right)
\frac{(s'-a)^{\frac{k}{2}-1}(2s'-3a)}{(s')^{\frac{3k}{2}+1}} .
\label{eq:ckhigh_integral}
\end{equation}
Using the partial-wave expansion and the definition of the positive
spectral average \eqref{eq:positiveaverage}, this expression can be
written compactly as
\begin{equation}
c_k^{\rm high}(a)
=
\bigl\langle \mathcal B_{k,a}(s',J) \bigr\rangle ,
\end{equation}
with kernel
\begin{equation}
\mathcal B_{k,a}(s',J)
=
\frac{P_J^{(D)}\!\left(
\sqrt{\frac{s'+3a}{s'-a}}
\right)}{(s')^{3k/2}}\,
(s'-a)^{\frac{k}{2}-1}\,
(2s'-3a).
\label{eq:Bka_def}
\end{equation}
Since $\rho_J(s')\ge0$, the sign and magnitude of $c_k^{\rm high}(a)$ are
entirely controlled by the kernel $\mathcal B_{k,a}$.

The low-energy contribution is fully determined within the effective
field theory and takes the form
\begin{equation}
-\,c_k^{\rm low}(a)
=
\left.
\frac{(-1)^{k/2}}{(k/2)!}\,
\partial_x^{k/2}\mathcal M(x,a)
\right|_{x=0}
-
\int_0^{M^2}
\frac{\mathrm ds'}{\pi}\,
\mathrm{Im}\,\mathcal M\!\left(s',\tau(s',a)\right)
\frac{(s'-a)^{\frac{k}{2}-1}(2s'-3a)}{(s')^{\frac{3k}{2}+1}} .
\label{eq:cklow}
\end{equation}
The first term captures the tree-level EFT contributions, while the
second encodes loop effects below the cutoff scale $M$.
In the following, we will focus on the tree-level analysis and neglect
the loop contribution for simplicity.


\subsubsection*{Explicit sum rules for scalar scattering with gravity}
\label{subsec:explicit_sumrules}

Restricting to tree-level scattering and neglecting loop effects,
the crossing-symmetric dispersive sum rules yield explicit relations
between low-energy EFT couplings and positive spectral averages.
For the first few even values of $k$, one finds
\begin{align}
k=2:\qquad
& -\frac{8\pi G}{a}
+ 2 g_2
- a\, g_3
=
\Biggl\langle
\frac{2s' - 3a}{(s')^{3}}\,
P_J^{(D)}\!\left(
\sqrt{\frac{s' + 3a}{s' - a}}
\right)
\Biggr\rangle
\label{eq:k2}, \\[0.4em]
k=4:\qquad
& 4 g_4
- 2 a\, g_5
+ a^{2} g_6
=
\Biggl\langle
\frac{(s' - a)(2s' - 3a)}{(s')^{6}}\,
P_J^{(D)}\!\left(
\sqrt{\frac{s' + 3a}{s' - a}}
\right)
\Biggr\rangle
\label{eq:k4}, \\[0.4em]
k=6:\qquad
& 8 g_6
- 4 a\, g_7
+ 2 a^{2} g_8
- a^{3} g_9
=
\Biggl\langle
\frac{(s' - a)^{2}(2s' - 3a)}{(s')^{9}}\,
P_J^{(D)}\!\left(
\sqrt{\frac{s' + 3a}{s' - a}}
\right)
\Biggr\rangle
\label{eq:k6}.
\end{align}

A notable advantage of these fully crossing-symmetric sum rules is that,
for fixed $k$, the low-energy side involves only a \emph{finite} set of
EFT couplings, without the need to construct linear combinations of
different sum rules.
As a result, these relations may be used directly in place of the
improved sum rules introduced in Ref.~\cite{Caron-Huot:2021rmr}.

\section{Sampling versus smearing}
\label{sec:distributions}

The dispersive sum rules derived above contain singular contributions
originating from the gravitational
$t$-channel pole. Because of this pole, the amplitude is
not well defined pointwise in the forward limit $t\to0$, or $a\to 0$.

The standard way to regulate this is to smear in $t$, as proposed in~\cite{Caron-Huot:2021rmr}. Instead of evaluating
the amplitude at a fixed kinematic point, or as Taylor series around a fixed point, one averages it against a test function,
\begin{equation}
\mathcal A_f \equiv \int \mathrm dt\, f(t)\,\mathcal A(t),
\end{equation}
with $f(t)$ smooth and compactly supported. This removes the singular
behavior at a point and makes the dispersive integrals finite. The
resulting sum rules are still linear in the EFT couplings, but now
involve smeared spectral averages of the form
\begin{equation}
\int \mathrm dt\, f(t)\,\langle \mathcal K_k(t,s',J)\rangle .
\end{equation}

The drawback is that smearing replaces a local problem in $t$ by a
nonlocal one. In particular, the simple pointwise linear constraints that
underlie  linearized unitarity can no
longer be imposed directly at each value of $t$, but can only be imposed after integration
against $f(t)$. This means one must also choose and
discretize a family of smearing functions and evaluate extra integrals,
which substantially complicates the optimization problem. For this
reason, while smearing regulates the graviton pole, it is not the most
convenient framework for all questions.

\subsection*{Sampling at finite resolution}
\label{subsec:functional}

The alternative approach we propose is to treat both the fixed-$t$ and the fixed-$a$ crossing-symmetric dispersive sum rules as \emph{functional relations} in the variable $t$, respectively $a$, to be probed at finite resolution rather than through smearing.
Concretely, the dispersive relation~\eqref{eq:fixeda_sumrule} is viewed as
an equality between ordinary functions of $t, a$, valid throughout the
interval
\begin{equation}
-1 < t < 0 ,\quad \textrm{or}\quad -\frac{1}{3} < a < 0,
\end{equation}

\paragraph{Fixed-$t$}
In our numerical implementation, we adopt the sampling points for $-1<t<0$
\begin{equation}\label{fixed-t}
t_i
=
-\frac{1}{2}
+\frac{1}{2}
\cos\!\left(\frac{i\pi}{N_t}\right),
\qquad i=1,\ldots,N_t-1,
\end{equation}
which cluster more densely near the endpoints of the interval $t \in [-1,0)$. Particular care is required near $t=-1$, and we therefore exclude this point in the numerical implementation. This is because the dispersion relation in Eq.~\ref{eq:B2imp} becomes singular at $\mu=1$, due to the denominator $\frac{1}{\mu^2 - t^2}$.\footnote{Numerically, this can render the linear optimization problem mildly ill-defined and require additional regularization.}

\paragraph{Fixed-$a$}
For fixed-$a$, with $-\frac{1}{3}<a<0$ we adopt a similar sampling method
\begin{equation}\label{fixed-a}
a_i
=
-\frac{1}{6}
+\frac{1}{6}
\cos\!\left(\frac{(2i-1)\pi}{2N_a}\right),
\qquad i=1,\ldots,N_a,
\end{equation}
with only the point $a=0$ requiring exclusion.

\subsection{Numerical implementation through primal bootstrap}
\label{sec:numerics}

To implement the dispersive sum rules numerically, we discretize the
high-energy spectral integrals and truncate the partial-wave expansion.
As a first step, we perform the change of variables
\begin{equation}
s' \equiv \mu = \frac{1}{z},
\qquad
z \in (0,1],
\label{eq 3.4}
\end{equation}
which maps the semi-infinite dispersive domain
$s' \in [M^2,\infty)$ to a finite interval in $z$.
The resulting integral is approximated by a Riemann sum with $N_\mu$
subdivisions, 
\begin{equation}
\sum_{J=0}^{\infty}
n^{(D)}_{J}
\int_{M^2}^{\infty}
\frac{\mathrm{d}s'}{\pi}\,
{s'}^{\,1-\frac{D}{2}}\,
\rho_J(s')\,(\cdots)
\;\longrightarrow\;
\sum_{\substack{0 \le J \le J_{\max} \\ J\;\mathrm{even}}}
n^{(D)}_{J}
\frac{1}{\pi N_\mu}
\sum_{i=1}^{N_\mu}
\left(\frac{i}{N_\mu}\right)^{\frac{D}{2}-3}
\rho\!\left(\frac{i}{N_\mu},J\right)
(\cdots),
\label{eq:riemann_discretization}
\end{equation}
where the power of $i/N$ arises from the Jacobian of the transformation
and the dimension-dependent dispersive measure.

We truncate the partial-wave expansion at a maximum spin $J_{\max}$ and
retain only even spins, as required by crossing symmetry.
The spectral density is therefore approximated by a finite set of
variables
\begin{equation}
\rho_J(s')
\;\longrightarrow\;
\rho\!\left(\frac{i}{N_\mu},J\right),
\qquad
1 \le i \le N_\mu,
\qquad
0 \le J \le J_{\max},
\end{equation}
which are subject to the linear unitarity bounds
\begin{equation}
0 \le \rho\!\left(\frac{i}{N_\mu},J\right) \le 2 .
\label{eq:rho_bounds}
\end{equation}

At fixed spacetime dimension $D$, the dispersive sum rules for
$k=2,4,6,\ldots$ are imposed as equality constraints at a discrete set of values
of the variable $t$. We sample the function at
a finite set of Chebyshev nodes.

With these approximations, the dispersive sum rules take the discretized
form
\begin{equation}
\sum_{\substack{0 \le J \le J_{\max} \\ J\;\mathrm{even}}}
\sum_{i=1}^{N_\mu}
n^{(D)}_{J}
\frac{1}{\pi N_\mu}
\left(\frac{i}{N_\mu}\right)^{\frac{D}{2}-3}
\rho\!\left(\frac{i}{N_\mu},J\right)
\,\mathcal{B}_k\!\left(t,\mu_i,J\right)
=
\mathcal{C}^{\mathrm{EFT}}_k(t),
\qquad
k=2,4,6,
\label{eq:discrete_sumrules}
\end{equation}
where $\mu_i \equiv N_\mu/i$ and the kernel $\mathcal{B}_k$ is given in \eqref{disp} for fixed-$t$, and 
\eqref{eq:Bka_def} for fixed-$a$.
The right-hand side $\mathcal{C}^{\mathrm{EFT}}_k(t)$ is a polynomial
function of $t$ whose coefficients are EFT couplings.

The numerical problem therefore reduces to a finite set of linear equality constraints
imposed on functions evaluated at discrete values of the kinematic variables, together
with positivity conditions on the discretized spectral density. All results presented in
this work are obtained by solving this truncated optimization problem and explicitly
verifying the stability of the bounds under systematic refinement of the ansatz
parameters. Further technical details are provided in Appendix~\ref{app:numerical_stability}.

\subsection{Numerical challenges and stability}

It should be emphasized that, despite its conceptual simplicity, the sampling approach is
numerically delicate. The main numerical issue is that the sampled functional constraints are highly sensitive to
how the truncations are taken. If the number of sampling points is increased while the
spin cutoff remains too low, the finite-spin ansatz cannot resolve the behavior required
near the graviton pole, and the constraints become effectively inconsistent. If, on the
other hand, the spin cutoff is increased too rapidly relative to the number of sampled
constraints, the optimization problem becomes underconstrained and the resulting bounds
develop an artificial drift. In practice, stable convergence is obtained only when the
kinematic sampling, spin truncation, and energy discretization are scaled together in a
carefully correlated way.

\begin{itemize}

\item \textbf{Chebyshev sampling}
The sampling points should be chosen as Chebyshev nodes in the relevant kinematic
interval, as done in \eqref{fixed-t}-(\ref{fixed-a}).
This choice clusters points near the endpoints, where the functional dependence is most
rapid, while automatically avoiding the singular points $t=0$ and $a=0$ associated with
the graviton pole. In the fixed-$t$ case it also avoids the endpoint $t=-1$, where the
kernel in Eq.~\eqref{disp} can develop an additional singularity.

\item \textbf{The sampling resolution and the spin cutoff must be scaled together.}
The number of sampling points and the maximal spin cannot be increased independently.
For fixed $N_t$ (or $N_a$), stable bounds are obtained only within a finite window
\[
r_{\min} N_t \;\lesssim\; J_{\max} \;\lesssim\; r_{\max} N_t,
\]
and analogously in the fixed-$a$ setup. In our fixed-$a$ analysis this window is numerically
close to
\[
2\,N_a^{(2)} \;\lesssim\; J_{\max} \;\lesssim\; 3\,N_a^{(2)}.
\]
The lower edge of this window has a clear physical origin: resolving the graviton pole, and
simultaneously satisfying an increasing number of null constraints, requires sufficiently
high-spin states. In the exact problem the pole itself requires an infinite tower of spins.
If $J_{\max}$ is chosen too small relative to the number of sampling points, the constraints
simply cannot be satisfied. The upper edge, by contrast, is a truncation artifact. Once
$J_{\max}$ is taken too large compared to the number of sampled constraints, the linear
program becomes effectively underconstrained, and the numerical bound begins to drift or
diverge.

\item \textbf{Higher-$k$ constraints must be introduced with a controlled hierarchy.}
The higher-$k$ dispersion relations can strengthen the bounds, but only if their sampling
resolution is kept compatible with the size of the ansatz. In particular, in the fixed-$a$
analysis we find that the $k=4$ constraints should be sampled more coarsely than the
$k=2$ ones. Concretely, stable results require roughly
\[
N_a^{(4)} \;\lesssim\; \frac{J_{\max}}{16}.
\]
If $N_a^{(4)}$ is taken too large at fixed $J_{\max}$, the system becomes unstable,
indicating that the ansatz space is too small to satisfy all imposed constraints
simultaneously. Thus, higher-order sum rules do not merely add information; they also
tighten the numerical requirements on the truncation.

\item \textbf{Energy discretization converges only after the previous conditions are met.}
Once the kinematic sampling and spin truncation are chosen inside the stable window, the
bounds converge rapidly with the number $N_\mu$ of points used to discretize
the dispersive integral. Outside the stable window, increasing the energy resolution does
not cure the instability and can even obscure the asymptotic behavior. In practice,
reliable extrapolation therefore requires correlated refinement of three ingredients at once:
the kinematic sampling, the spin cutoff, and the energy discretization.

\item \textbf{Fixed-$t$ remains fragile, while fixed-$a$ is robust.}
Even after all of the above improvements, the fixed-$t$ formulation still converges rather
slowly and remains noticeably more susceptible to numerical instabilities, especially for
non-projective bounds. By contrast, the manifestly crossing-symmetric fixed-$a$
dispersion relations are substantially more stable, converge faster, and allow us to extract
non-projective bounds together with highly stable extremal spectra. This is the main
practical reason why our final non-projective results are obtained in the fixed-$a$
framework.

\end{itemize}

Sampling is indeed the most direct way to
treat the graviton pole, but it becomes viable only after the truncations are correlated in
a careful way. Once this is done, the method reproduces the known projective bounds,
yields new non-projective bounds, and gives direct access to the corresponding extremal
spectra. 

\section{Results without gravity}
\label{sec:4d_nogravity}
In this section we set $8\pi G=0$, i.e.\ we switch off gravitational
interactions, and fix the dispersive mass scale to $M=1$. We use this simpler
setup to test and illustrate the sampling method. Rather than starting from the
fixed-$t$ dispersion relations, we work directly with the crossing-symmetric
fixed-$a$ representation. 

Our aim here is primarily methodological rather than the derivation of
provably optimal bounds. We wish to show concretely how the \emph{functional}
implementation of crossing-symmetric dispersive sum rules can be performed in a
finite-resolution sampling framework. We find that the bounds obtained from
positivity alone already lie numerically close to the known optimal bounds.
Once linearized unitarity is imposed, the resulting non-projective bounds are stronger than those found in earlier analyses \cite{Caron-Huot:2020cmc,Chiang:2022ltp}, reflecting the fact that crossing-symmetric dispersion relations effectively encode an infinite number of null constraints.

These non-projective bounds should, however, be interpreted with some care:
linearized unitarity is a controlled approximation only in the weakly
coupled regime, and the bounds we obtain may extend beyond the domain
where that approximation is quantitatively reliable.

We begin by illustrating the primal approach in the $a\to0$ limit, equivalently in the forward limit, where the structure of the constraints is simplest.  We then generalize the analysis by sampling away from $a\to0$, which allows us to test the method beyond the forward limit.

\subsection{Expansion around $a=0$}
We illustrate the construction in $D=4$ starting from the $k=2$ kernel
\begin{equation}
\mathcal{B}_2(a,\mu,J)
=
\frac{2\mu-3a}{\mu^3}\,
P_J^{(4)}\!\left(
\sqrt{\frac{\mu+3a}{\mu-a}}
\right).
\end{equation}

Assuming the kernel is regular in the limit $a\to0$, one may expand it around $a=0$ and match each derivative to the low-energy EFT amplitude. The first nontrivial terms reproduce the usual dispersive sum rules,
\begin{equation}
\left\langle \frac{2}{\mu^3} \right\rangle = 2g_2,
\qquad
\left\langle \frac{-3+2J(J+1)}{\mu^4} \right\rangle = -g_3,
\end{equation}
so positivity implies $g_2>0$. Starting at sufficiently high order in the $a$-expansion, however, there is no independent tree-level EFT coefficient left to match. The corresponding derivatives must therefore vanish, producing null constraints. For the first such case one finds
\begin{equation}
\left\langle
\frac{J(J+1)\bigl(J^2+J-8\bigr)}{\mu^5}
\right\rangle
=0.
\end{equation}
In this way, crossing symmetry generates an infinite tower of linear relations on the spectral data.

We next follow the discussion in Section~\ref{sec:numerics}: we truncate the spin, discretize the positive spectral density $\rho_{l,i}$, and treat the sum rules as linear functionals of $\rho$. Fixing the overall normalization by $g_2[\rho]=1$, imposing positivity $\rho_{l,i}\ge 0$, and adding the null constraint $\mathcal{N}_4[\rho]=0$, the problem becomes a linear program minimizing $g_3[\rho]$. 

Using a discretization with $N_\mu = 400$ points and a spin truncation $J_{\max} = 40$, we obtain the bound
\begin{equation}
g_3^{\min} = -10.6125.
\end{equation}

This agrees with the dual formulation and provides a simple proof of principle that null constraints can be incorporated efficiently in a primal optimization framework.

\subsection{Sampling at finite $a$ resolution}
\label{subsec:functional_crossing}

We now reformulate the dispersive analysis in a \emph{functional}
language by imposing crossing symmetry through its evaluation at several
kinematic points, rather than through a Taylor expansion around
$a=0$. Instead of matching the sum rule order by order in derivatives at
a single point, we require it to hold at a finite set of values of the crossing parameter $a$ within its domain of
validity. 

Concretely, we discretize the spectral integral and the partial-wave sum
and impose the $k=2$ dispersive relation
\begin{equation}
\mathcal{S}(2,a,d)
\equiv
\sum_{J,z}
n_{J,d}\,
\frac{1}{\pi}\,
\frac{1}{N_\mu}\,
\left(\frac{z}{N_\mu}\right)^{1-\frac{d}{2}}
\rho(z,J)\,
B(2,a,N_\mu/z,J,4)
=
2 g_2 - g_3\, a\,,
\label{eq:S2_def}
\end{equation}
at a discrete set of values $a=a_i$.
Enforcing the equality at multiple values of $a$ should be viewed as a
finite-resolution approximation to a functional identity in $a$.

The sampling points $a_i$ are chosen according to a Chebyshev-type
distribution,
\begin{equation}
a_i
=
-\frac{1}{6}
+\frac{1}{6}
\cos\!\left(\frac{(2i-1)\pi}{2N_a}\right),
\qquad i=1,\ldots,N_a,
\end{equation}
which efficiently probes the interval while improving conditioning of
the resulting linear system.

The resulting problem can again be formulated as a linear program. One
minimizes $g_3$ subject to positivity of the spectral density and the
crossing-symmetric dispersive constraints evaluated on a finite set of
sampling points in $a$. After discretization, the optimization variables
are the spectral densities together with the relevant EFT couplings, and
all constraints are linear. We relegate the detailed discretization and
sampling procedure to the appendix.

For $D=4$, with our default numerical settings, using just the $k=2$ sum rule, we find
\begin{equation}\label{bound1}
g_3^{\min} = -10.3604.
\end{equation}
This bound is stronger than the one obtained from imposing only a single
null constraint at $a=0$. The improvement comes from enforcing the
$k=2$ dispersive sum rule over an interval in $a$, which effectively
captures a large subset of the null constraints encoded in the
crossing-symmetric dispersion relation.

Linearized unitarity can be incorporated straightforwardly in the same
framework by imposing the additional upper bound on the spectral density.
More general optimizations involving higher-$k$ sum rules can be treated
analogously; we will use this setup repeatedly below, and leave the full
linear-programming formulation to the appendix.

We next add the constraints from higher $k$ sum rules, $k=4,6$, from \eqref{eq:k4} and (\ref{eq:k6}). Unless otherwise stated, our numerical results are obtained
with
\begin{equation}
N_\mu=300,\qquad J_{\max}=40,\qquad
N_a^{(2)}=N_a^{(4)}=N_a^{(6)}=20,
\end{equation}
We have checked that increasing these truncation parameters does not
lead to appreciable changes in the bounds.

\paragraph{Results for projective $g_4/g_2$ and $g_3/g_2$ in four dimensions.}
\FloatBarrier
\begin{figure}[t]
\centering
\includegraphics[width=0.8\textwidth]{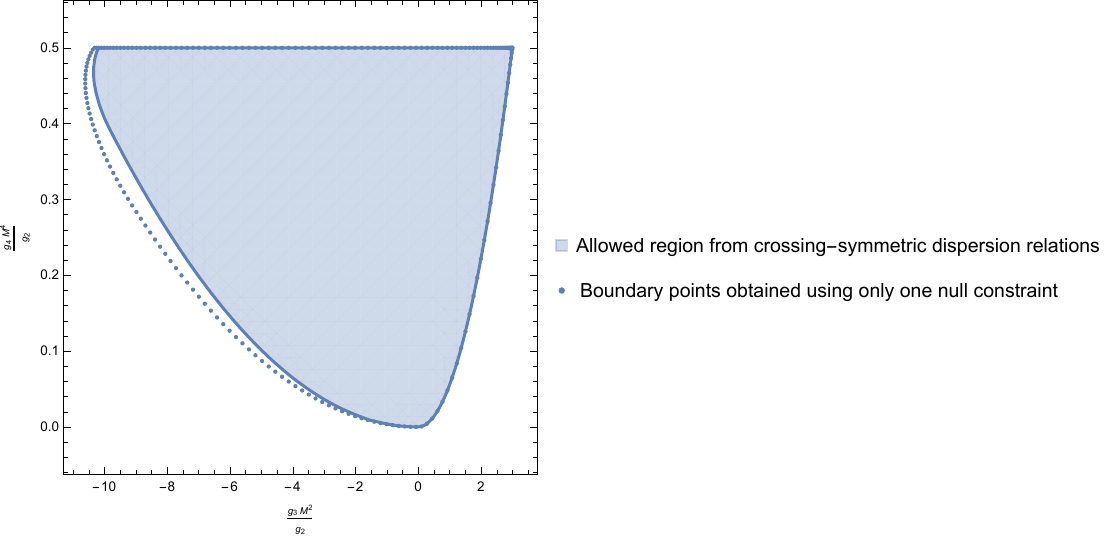}
\caption{
Allowed region in the $(g_3/g_2,\, g_4/g_2)$ plane derived from
crossing-symmetric dispersive sum rules with finite sampling. The
result is in excellent agreement with Ref.~\cite{Caron-Huot:2020cmc},
which used the dual bootstrap with a large number of null constraints.
For comparison, we also show the boundary obtained in the forward limit
from imposing only a single null constraint.}
\label{fig:g3g4_comparison}
\end{figure}

Enforcing the higher $k$ sum rules strengthens the bound on $g_3$ to
\begin{equation}
g_3^{\min} \simeq -10.3475,
\end{equation}
which is close to the optimal projective value
$g_3^{\min} \simeq -10.3465$ reported in
Ref.~\cite{Caron-Huot:2020cmc}.

The remaining small discrepancy with Ref.~\cite{Caron-Huot:2020cmc} can
be attributed to the finite number of sum rules and sampling points used
in our analysis. In principle, this gap can be reduced systematically by
including additional crossing-symmetric sum rules at higher $k$ and/or a
denser sampling in $a$.
\paragraph{Results for non-projective $g_2$ and $g_3$ in four dimensions.}

\begin{figure}[t]
\centering
\includegraphics[width=1\textwidth]{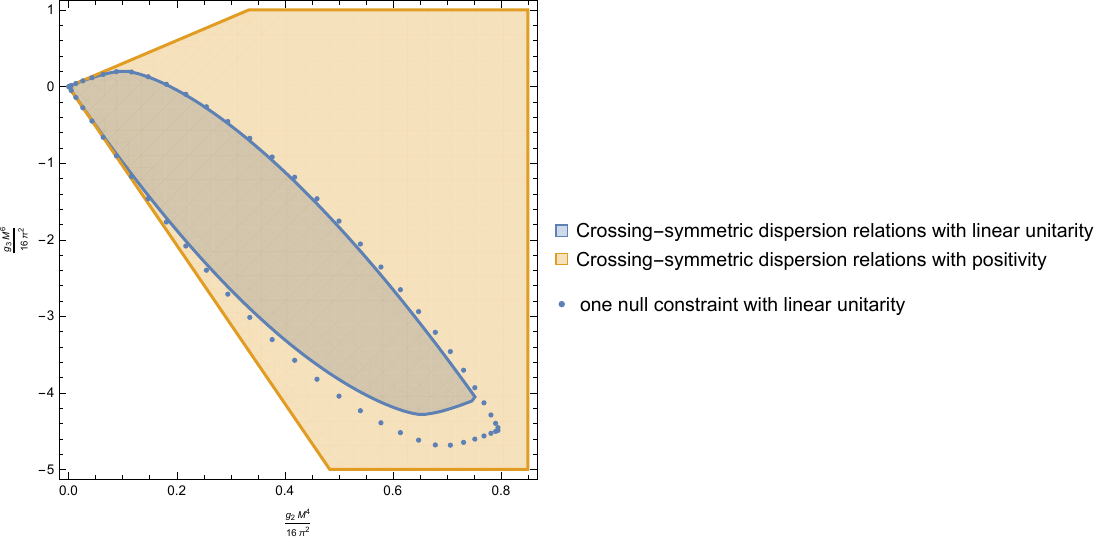}
\caption{
Allowed region in the $(g_2,\, g_3)$ plane derived from
crossing-symmetric dispersive sum rules.
We show results obtained with positivity alone and with the optional
inclusion of linearized unitarity, together with the boundary obtained
from imposing a single tree-level null constraint for comparison.}
\label{fig:g3g2_comparison}
\end{figure}
We now impose the unitarity upper bound $\rho\le 2$, and leave $g_2$ as a free coupling. We find the upper bound on $g_2$ obtained using a single null constraint agrees
remarkably well with Ref.~\cite{Caron-Huot:2020cmc},
\begin{equation}
\frac{g_2^{\max}}{(4\pi)^2} < 0.79.
\end{equation}
In that work, the spectral density was separated into low-spin and
high-spin contributions, and it was shown that both sectors yield
convergent bounds once linearized unitarity is imposed. The resulting
constraint can be solved analytically by reformulating the problem as a
double $L$-moment problem and intersecting with the null surface, as explained in Ref.~\cite{Chiang:2022jep,Chiang:2022ltp}

By contrast, imposing the crossing-symmetric dispersive relation at
multiple values of $a$ strengthens the bound to
\begin{equation}
\frac{g_2^{\max}}{(4\pi)^2} < 0.75,
\end{equation}
reflecting the additional kinematic information captured by enforcing
crossing symmetry over an interval in $a$. This improvement is visible
in Fig.~\ref{fig:g3g2_comparison}.

For the two-dimensional allowed region in the $(g_2,g_3)$ plane, we find
that incorporating linearized unitarity leads to stronger bounds than
those obtained in Ref.~\cite{Chiang:2022jep,Chiang:2022ltp}, where the problem was
formulated as a three-dimensional Minkowski sum, but only included one null constraint.

If one allows fewer subtractions in the dispersive representation, even
stronger bounds can be obtained using the $\rho$-bootstrap approach of
Ref.~\cite{Paulos:2017fhb}, which also allows one to
impose full unitarity constraints, as explored in
Ref.~\cite{Chen:2022nym}, leading to a significant further reduction
of the allowed parameter space. The use of fewer subtractions, however, requires extra
ultraviolet assumptions. This issue was emphasized in Ref.~\cite{EliasMiro:2022xaa},
and motivates our focus on twice-subtracted, crossing-symmetric dispersive sum rules. However, our approach can also easily accommodate stronger UV behavior and therefore fewer subtractions.

\section{Bounds and extremal spectra with gravity}
\label{sec:5d6d_gravity}

In this section we turn to the case with dynamical gravity. We first 
use the fixed-$t$ improved sum rules as a benchmark for the primal
sampling approach at the projective level, where comparison with the smeared
analysis is most direct.  We then use the fully crossing-symmetric fixed-$a$
dispersion relations to derive non-projective bounds
and extract extremal spectra under linearized unitarity,
\[
0 \le \rho \le 2.
\]
This division of frameworks is motivated by numerics: fixed-$a$ is substantially
more stable for the non-projective problem.

We present convergence tests in Appendix~\ref{app:numerical_stability}.  We briefly comment on loop effects in Appendix~\ref{loopchecks} and leave a systematic treatment to future work.

\subsection{Projective bounds}
\label{subsec:proj_5d6d}
\FloatBarrier

\subsubsection*{Fixed-$t$}
\begin{figure}[htbp]
    \centering
        \includegraphics[height=8.cm]{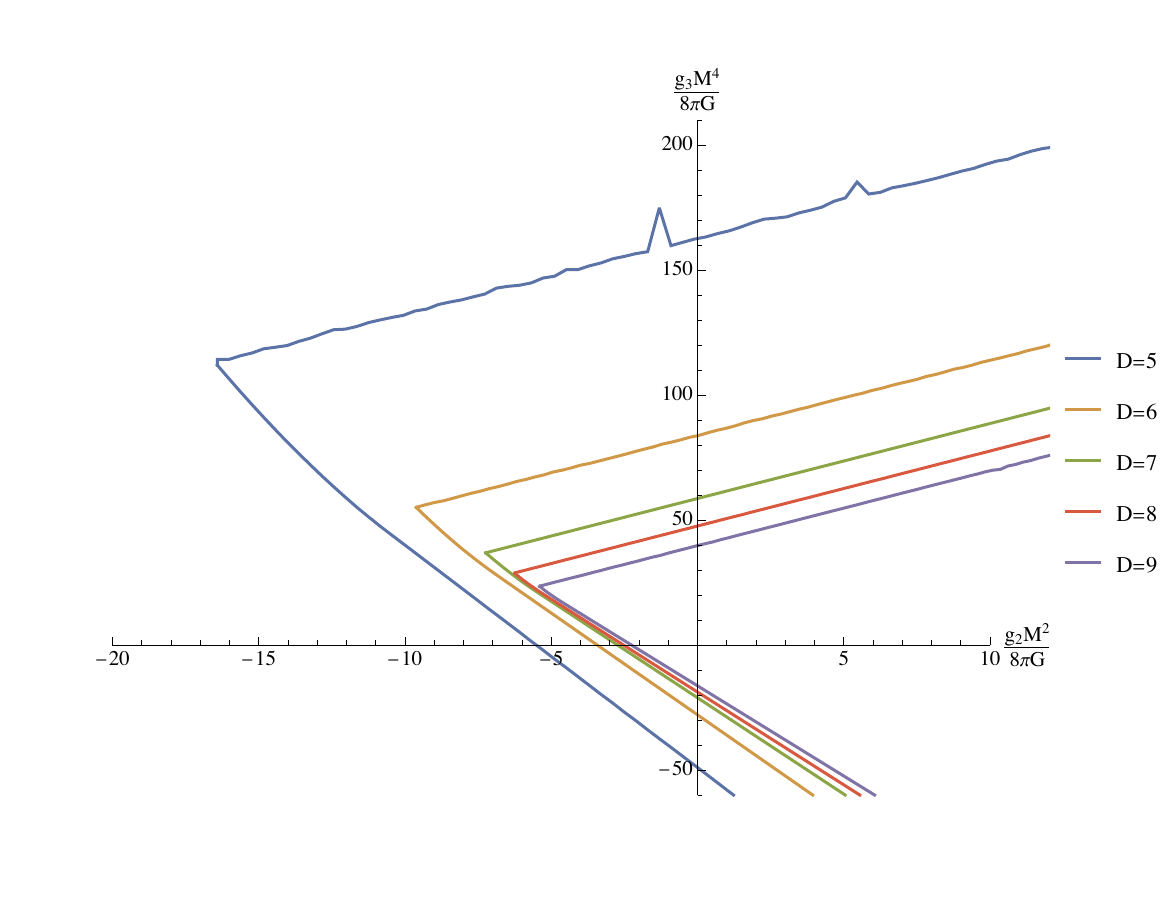}
        \caption{Allowed regions in ($g_2/8\pi G,g_3/8\pi G$) plane obtained via the primal sampling method. For \(D = 5\), we set \(N_t = 60\), \(N_\mu = 600\), and \(J_{\max} = 1200\). For these high parameter values, the upper boundary exhibits minor numerical instabilities; for more details, see Appendix \ref{A.3}. For other dimensions, we set \(N_t = 25\), \(N_\mu = 400\), and \(J_{\max} = 300\). Larger parameter values only slightly improve the results in $D\ge 6$. Except $D=5$, the bounds agree well with those in Ref.~\cite{Caron-Huot:2021rmr}, Figure 4.}
    \label{projective_bound_multi_dimensions}
\end{figure}

Overall, our results, shown in Fig.\ref{projective_bound_multi_dimensions}, are in broad agreement with the
projective bounds obtained in Ref.~\cite{Caron-Huot:2021rmr}. The main exception
is $D=5$, where we find a somewhat larger deviation. More explicitly, our lower
bounds read
\begin{align}
\frac{g_2}{8\pi G} &\gtrsim -16.5 \qquad (D=5), \label{eq:D5projectivebound}\\
\frac{g_2}{8\pi G} &\gtrsim -9.77 \qquad (D=6). \label{eq:D6projectivebound}
\end{align}
For comparison, the corresponding smeared values are approximately
$-18$ in $D=5$ and $-9.57$ in $D=6$. Thus our sampled primal approach yields a
slightly stronger bound in $D=5$ and a slightly weaker one in $D=6$.

\subsection{Projective extremal spectra} 
An attractive feature of the present functional formulation is that it provides direct access to the spectrum of extremal solutions. Unlike dual or smearing-based approaches, where extracting spectral information typically requires solving an auxiliary optimization problem, the primal linear program yields the extremal spectral density $\rho_J(\mu)$ explicitly as part of the solution.

\begin{figure}[t]
\centering
\includegraphics[width=1\textwidth]{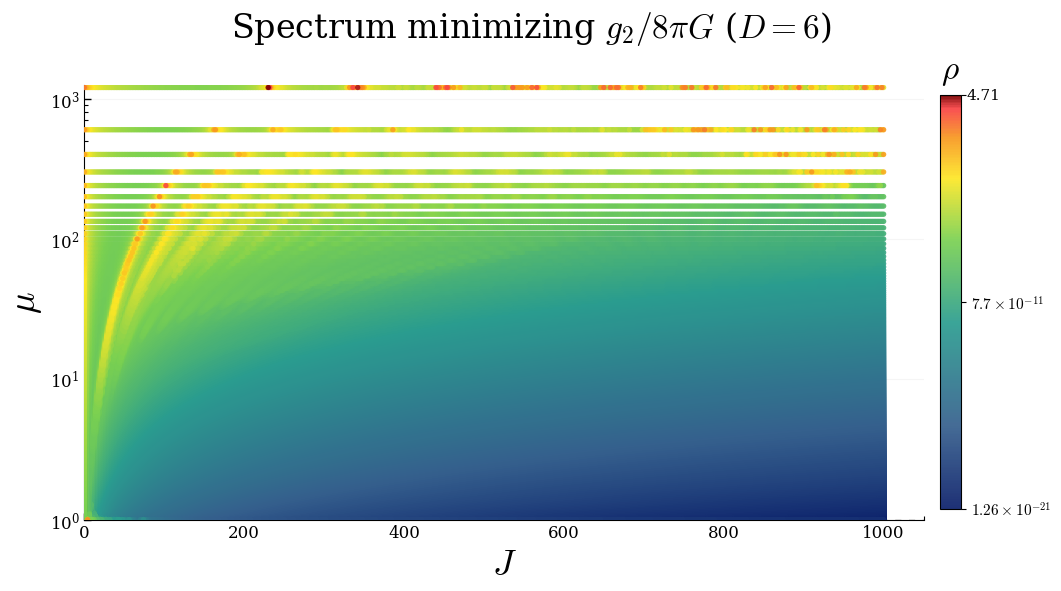}
\caption{Log plot of
extremal spectral density $\rho_J(\mu)$ in $D=6$ at the projective lower bound of $g_2/8\pi G$. The extremal spectrum shows quadratic Regge-like trajectories in the $(J,\mu)$ plane. Away from these trajectories, the spectral density is strongly suppressed. The result is obtained with $N_t=60$, $J_{\max}=1000$, and $N_\mu=1200$.}
\label{fig:extremal_spectrum_5d_intro}
\end{figure}

In Fig.~\ref{fig:extremal_spectrum_5d_intro} we plot the extremal spectral density for the lower bound of $g_2/8\pi G$ in $D=6$. Interestingly, although the resulting bound is close to that obtained in Ref.~\cite{Caron-Huot:2021rmr} using the dual bootstrap ($g_2/8\pi G\ge -9.77$ vs. $g_2/8\pi G\ge-9.57$), the corresponding extremal spectrum is qualitatively very different. In the primal bootstrap, states appear to populate the full allowed region, but the vast majority carry parametrically smaller spectral weight $\rho$. The dominant support organizes into quadratic Regge-like trajectories in the $(J,\mu)$=(spin,mass) plane, while away from these trajectories the spectral density is strongly suppressed. By contrast, the extremal spectrum found in Ref.~\cite{Caron-Huot:2021rmr} populated only the lower-right triangular region, with clearly separated states.

\subsubsection*{Fixed-$a$}

\begin{figure}[t]
\centering
\includegraphics[width=0.6\textwidth]{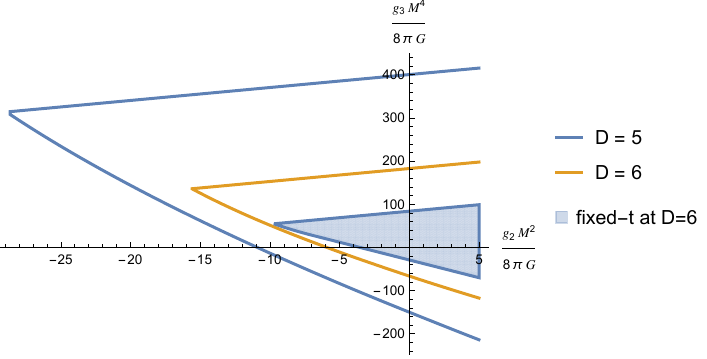}
\caption{
Numerical projective bounds on EFT coefficients with positivity condition using fixed-$a$ dispersion relations. 
For comparison, we also include the result in $D=6$ from using fixed-$t$.
These bounds are numerically close to those obtained from smeared fixed-$a$
sum rules~\cite{Beadle:2025cdx,Chang:2025cxc}, but are weaker than the
bounds derived from improved fixed-$t$ sum rules~\cite{Caron-Huot:2021rmr}. 
}
\label{fig:g3g2_gravity}
\end{figure}

Using sampling with fixed-$a$, we find projective bounds that are numerically
close to those obtained from smeared fixed-$a$ dispersion relations
\cite{Beadle:2025cdx,Chang:2025cxc,Pasiecznik:2025eqc}. We show the corresponding
allowed regions in the $(g_2/8\pi G, g_3/8\pi G)$ plane for $D=5,6$ in Fig.~\ref{fig:g3g2_gravity}.
These bounds are, however, weaker than those obtained from the improved fixed-$t$
sum rules. This is consistent with the role of fixed-$a$ in our analysis:
fixed-$a$ provides the numerically stable framework for the non-projective
problem, rather than the sharpest projective benchmark.

\subsection{Non-projective bounds}
\label{subsec:5d_nonprojective}
\subsubsection*{Fixed-$t$}

In our present implementation, the fixed-$t$ dispersion relations do not lead to stable or convergent non-projective bounds. The most likely reason is that non-projective bounds are significantly more sensitive to higher-order null constraints, while in the fixed-$t$ approach the inclusion of additional constraints tends to trigger numerical instabilities before convergence is reached. It is possible that a more refined treatment---for example in the choice of sampling points, the discretization of the energy integral, or the spin truncation---could improve the situation.

For this reason, in the remainder of this work we instead use the fixed-$a$ dispersion relations to derive non-projective bounds.

\subsubsection*{Fixed-$a$}
We now focus on $D=5$ to illustrate how the bounds are modified once
linearized unitarity is imposed. In Ref.~\cite{Chiang:2022ltp},
linearized unitarity was implemented via a Minkowski-sum construction,
which geometrically deprojects the EFT coupling space. That approach,
however, is obstructed in the presence of a gravitational pole: the
infrared divergence associated with massless graviton exchange prevents
a straightforward application of the Minkowski-sum method.

\paragraph{Bounds on the gravitational coupling.}

We begin by extracting upper and lower bounds on the gravitational
coupling. In five spacetime dimensions, our numerical analysis yields
\begin{equation}
0 \;<\; \frac{8\pi G}{(4\pi)^{5/2}}\, M^{3} \;<\; 0.86 ,
\qquad (D=5) .
\end{equation}
The lower bound simply reflects the fact that consistent solutions
require gravity to be attractive, as already implied by the positivity
of the spectral density and the small-$a$ behavior of the dispersive sum
rules. The upper bound is more interesting and admits a natural physical
interpretation.

Writing $8\pi G = M_P^{-3}$ in five dimensions, the bound becomes
\begin{equation}\label{npbound}
0 \;<\; \frac{M}{M_P} \;\lesssim\; 7.8 \, .
\end{equation}
Thus the EFT cutoff cannot be parametrically separated from the Planck scale.

Equivalently, the bound reflects the intrinsic limitation of EFT coupled to gravity: as $M \to M_P$, the dimensionless coupling $8\pi G M^3$ becomes $\mathcal{O}(1)$, and the low-energy expansion breaks down. Our analysis shows that this breakdown is enforced by analyticity, crossing symmetry, positivity, and linearized unitarity.

Importantly, the result is model-independent. It does not rely on assumptions about the ultraviolet completion, spectrum, or symmetries, and follows solely from universal properties of relativistic quantum field theory combined with the long-range nature of gravity.

\paragraph{Non-projective bounds on $(8\pi G, g_2)$-space.}
\FloatBarrier

\begin{figure}[t]
\centering
\begin{overpic}[width=0.6\textwidth]{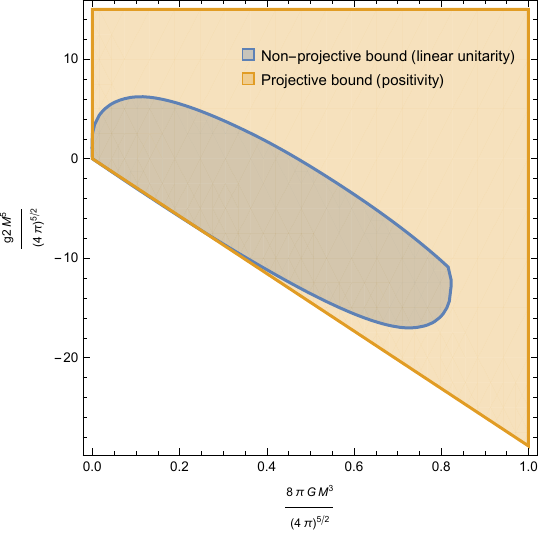}

    \put(6,66){\color{red}\linethickness{0.8pt}\framebox(18,13){}}

    \put(17,30){
        \color{black}
        \setlength{\fboxsep}{2pt}
        \fcolorbox{red}{white}{   
            \begin{minipage}{0.25\textwidth}
                \includegraphics[width=\linewidth]{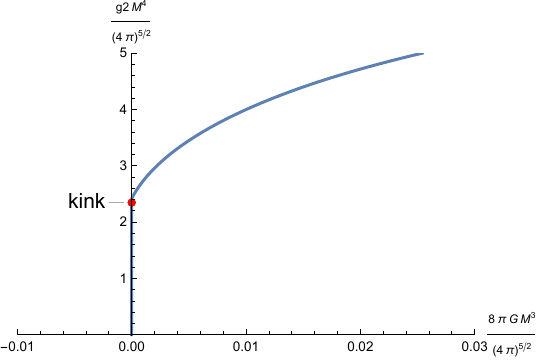}
            \end{minipage}
        }
    }

\end{overpic}

\caption{
Allowed regions in the $(8\pi G,\, g_2)$ plane for a scalar coupled to
gravity in flat spacetime in $D=5$ (with heavy mass scale $M$).
We compare positivity-only bounds with bounds including linearized
unitarity. Imposing linearized unitarity renders the allowed region
compact in both $8\pi G$ and $g_2$. The red box highlights the region
near $8\pi G \simeq 0$ that is magnified in the inset. The zoom reveals
a kink in the upper boundary, indicating that the $G \to 0$ limit
is non-smooth.}
\label{fig:g2G_5D}
\end{figure}

For small values of \(8\pi G\), the positivity-only and linearized-unitarity bounds are in good agreement. Once linearized unitarity is imposed, however, both an upper and a lower bound on \(g_2\) appear, as shown in Fig.~\ref{fig:g2G_5D}. The limit \(8\pi G \to 0\) is not smooth, since $g_2$ has an upper bound even when $8\pi G=0$.

\begin{figure}[t]
\centering
\includegraphics[width=0.7\textwidth]{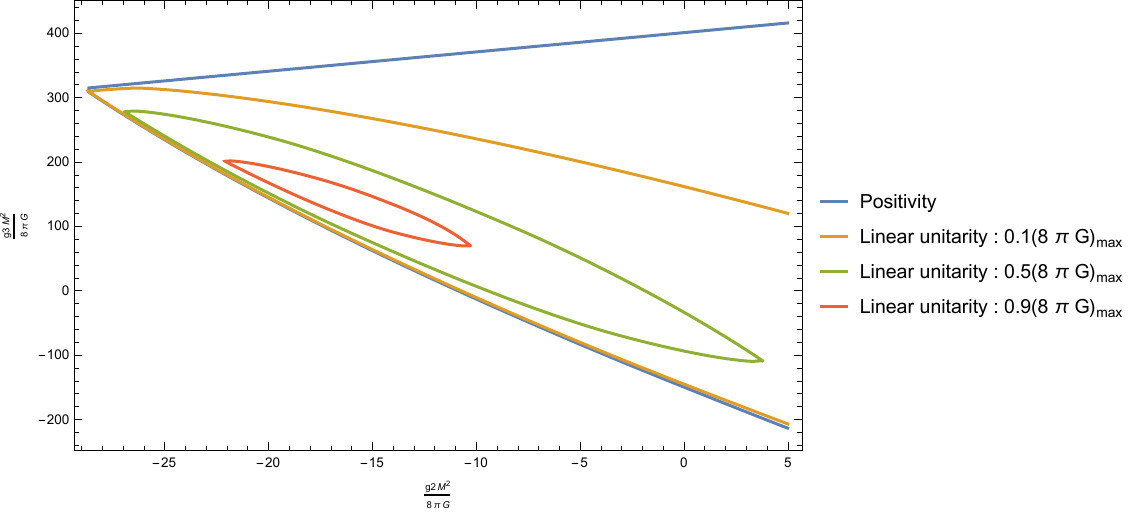}
\caption{
Comparison of projective (positivity-only) and non-projective (including
linearized unitarity) bounds in five dimensions for several fixed values
of the gravitational coupling,
$8\pi G = 0.1,\; 0.5,$ and $0.9$ times its maximal allowed value,
$(8\pi G)_{\max} = 0.86\,(4\pi)^{5/2}/M^3$.
As $8\pi G$ increases, the allowed region deforms smoothly.}
\label{fig:seriesG_5D}
\end{figure}

We finally compare projective (positivity-only) bounds with
non-projective bounds (including linearized unitarity) at fixed
gravitational coupling. We consider representative values
 $8\pi G = 0.1,\; 0.5,\; \text{and}\; 0.9 \times (8\pi G)_{\max}$ and plot the corresponding allowed regions in the $(g_2,g_3)$ plane, shown in Figure \ref{fig:seriesG_5D}.

\subsection{Non-projective extremal spectra} \label{sec:extremal_spectrum}

For the non-projective bounds found in the previous section our primal bootstrap approach returns a sharply localized spectral distribution in the $(J,\mu)$ plane, where $J$ denotes the angular momentum and $\mu$ the dispersive mass variable.

\begin{figure}[t]
\centering
\includegraphics[width=0.85\textwidth]{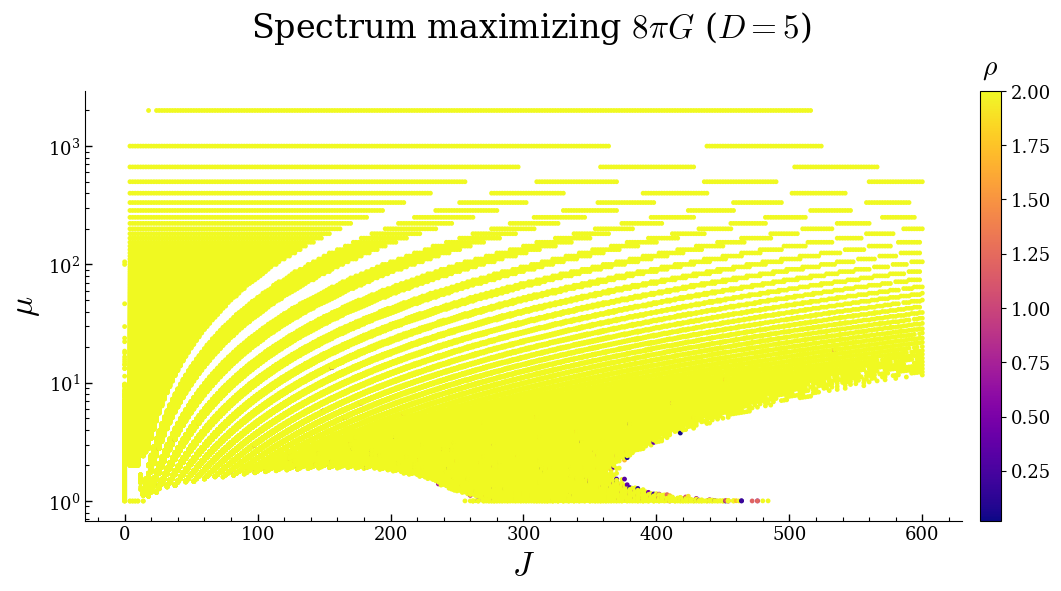}
\caption{Log plot for the 
extremal spectral density $\rho_J(\mu)$ in $D=5$ at the maximal allowed
gravitational coupling  $\frac{M}{M_P}\le 7.8$.
The extremal solution has a sharp structure ($\rho$ is mostly either exactly 0 or 2) and organizes into multiple
quadratic Regge-like bands in the $(J,\mu)$ plane. The result is obtained from a numerical solution with $N_a^{(2)}=200$, $J_{\max}=600$, $N_\mu=2000$. In $D=5$, the empty regions of space are sensitive to truncation.}
\label{fig:extremal_spectrum_5d}
\end{figure}

\begin{figure}[t]
\centering
\includegraphics[width=0.7\textwidth]{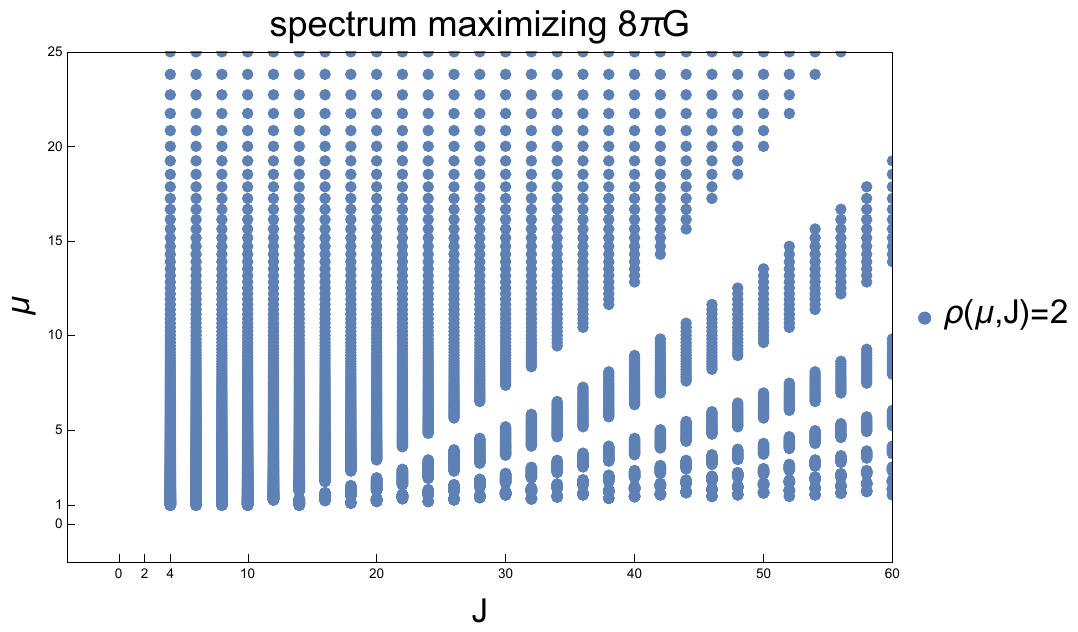}
\caption{
Extremal spectral density $\rho_J(\mu)$ in $D=5$ at the maximal allowed
gravitational coupling  $\frac{M}{M_P}\le 7.8$.
The extremal solution has a sharp structure ($\rho$ is mostly either exactly 0 or 2) and organizes into multiple
quadratic Regge-like bands in the $(J,\mu)$ plane. The picture shows a zoom-in of a spectrum  obtained for $N_t=200, J_\textrm{max}=400$, $N_\mu=500$.}
\label{fig:extremal_spectrum_5d_zoom}
\end{figure}

Fig.~\ref{fig:extremal_spectrum_5d} shows a representative example of the extremal spectral density obtained in $D=5$ at the maximal allowed gravitational coupling in eq.\eqref{npbound}. The dominant band structure is stable under variation of the ansatz parameters, while some of the empty regions are more sensitive to truncation. A zoom into the small-spin, small-mass region is shown in Fig.~\ref{fig:extremal_spectrum_5d_zoom}.

\begin{figure}[t]
\centering
\includegraphics[width=0.9\textwidth]{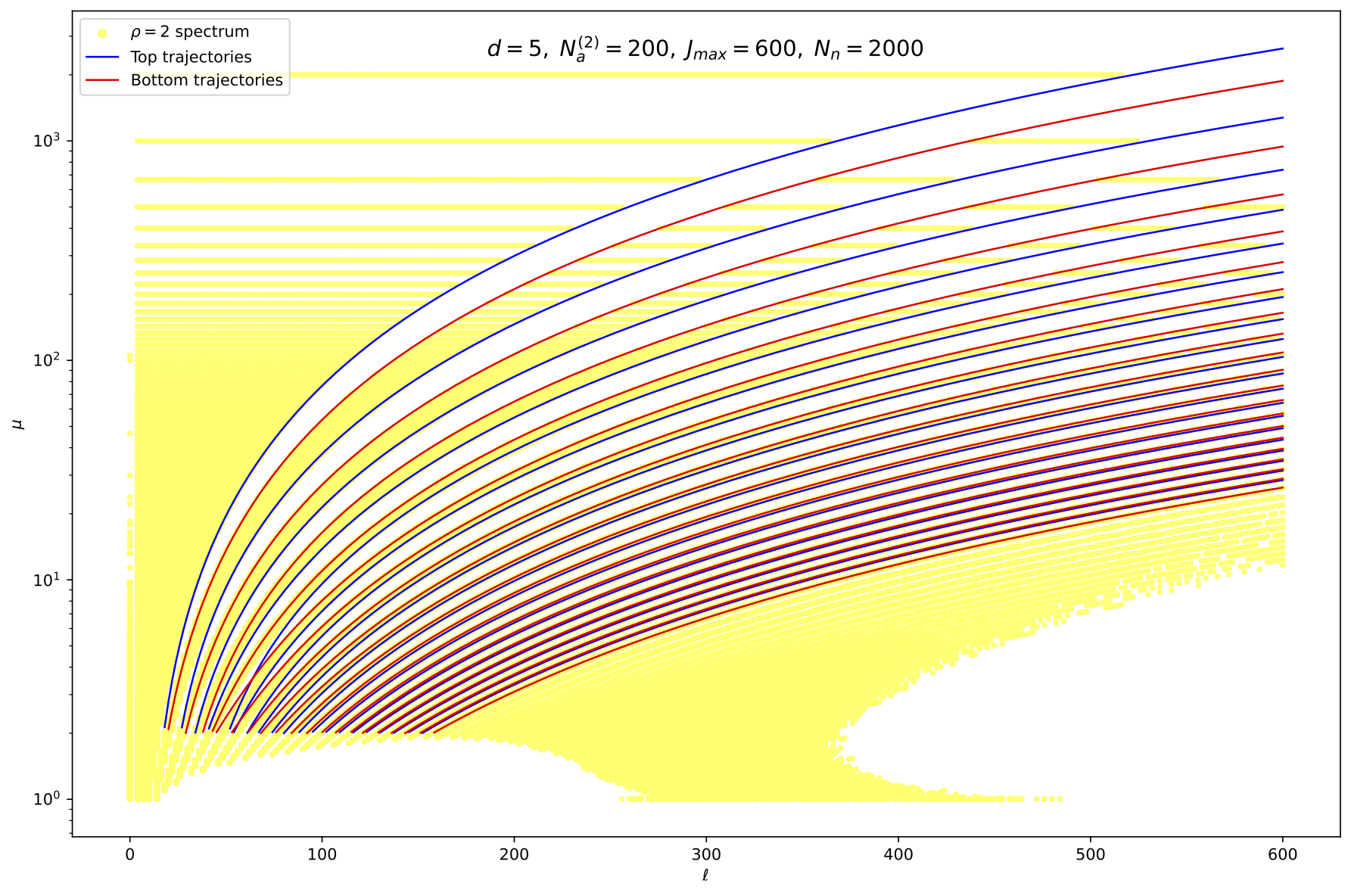}
\caption{
The spectrum exhibits bands that are bounded by quadratic trajectories of the type $\mu_n=b_{0;n}+b_{1;n} J+b_{2;n}J^2$.
}
\label{fig:extremal_spectrum_5d_traj}
\end{figure}

Remarkably, as shown in Figure \ref{fig:extremal_spectrum_5d_traj} the spectrum is characterized by narrow bands of support, whose boundaries (which we will refer to bottom and top) are extremely well approximated by quadratic relations between spin and mass,
\begin{equation}
\mu_n\sim b_{0,n}+b_{1,n} J+ b_{2,n}J^2\,,
\end{equation}
reminiscent of Regge trajectories.

Linear trajectories are a familiar feature of high-energy scattering and are often associated with string-like behavior, where an infinite tower of higher-spin states lies on approximately linear trajectories. However, the quadratic trajectories we uncover are a novel feature, and it would be interesting to understand what theories possess such spectra. 

Furthermore, the leading coefficients $b_{2,n}$ themselves follow a simple inverse-quadratic
pattern in the band number,
\begin{equation}
b_{2,n}\sim \frac{A}{(B+n)^2}\, ,
\label{eq:b2fit}
\end{equation}
with different constants $A$ and $B$ for the upper and lower boundaries. For example in
$D=5$ we find
\begin{equation}
b_{2,n}=\frac{0.0364}{(1.243+n)^2}
\qquad \text{for the top trajectories,}
\label{eq:b2top}
\end{equation}
and
\begin{equation}
b_{2,n}=\frac{0.0318}{(1.481+n)^2}
\qquad \text{for the bottom trajectories.}
\label{eq:b2bottom}
\end{equation}

\begin{figure}[t]
\centering
\includegraphics[width=0.8\textwidth]{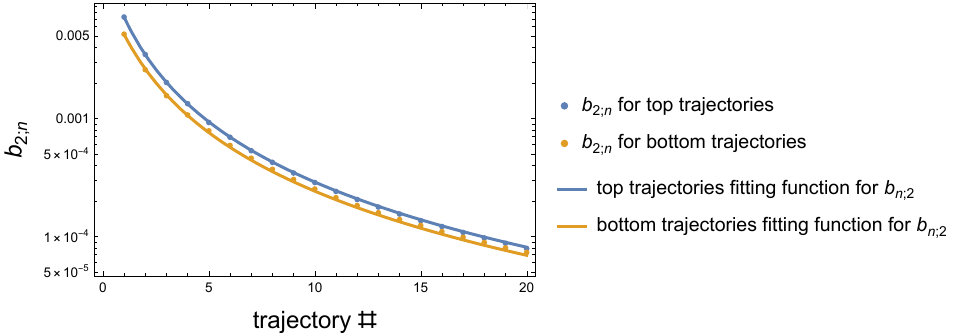}
\caption{
The leading coefficients $b_{2;n}$ of the trajectories $n$ of the type $\mu_n=b_{0;n}+b_{1;n} J+b_{2;n}J^2$ themselves follow an inverse quadratic fitting function $b_{2;n}=\frac{A}{(B+n)^2}$, with $A,B$ different for the top and bottom trajectories.
}
\label{fig:c2_fit}
\end{figure}

These fits, shown in Fig.~\ref{fig:c2_fit}, provide a compact characterization of the stable part of the extremal spectrum.

\begin{figure}[ht]
\centering
\begin{subfigure}{0.48\textwidth}
\centering
\includegraphics[width=\linewidth]{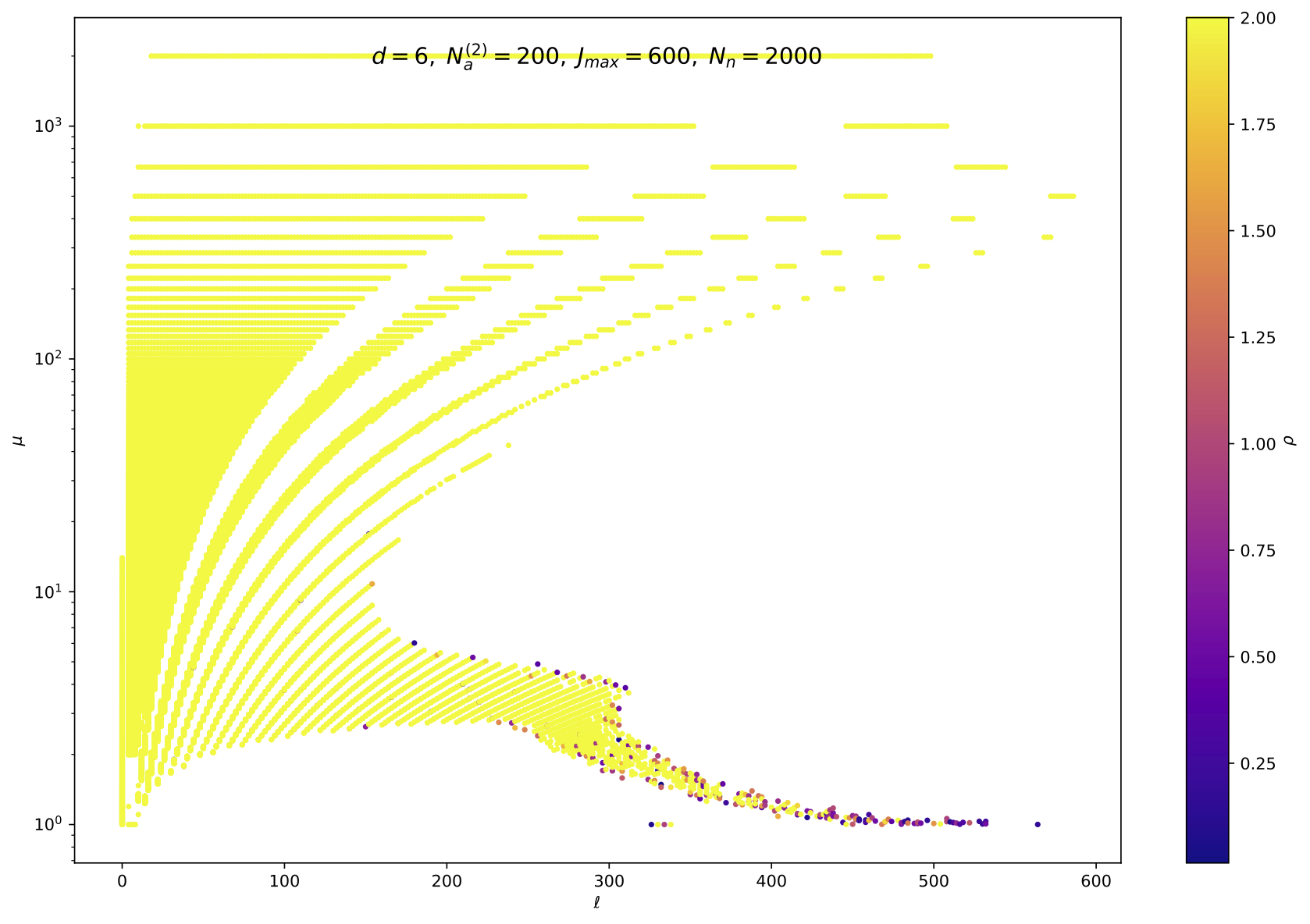}
\caption{$D=6$}
\end{subfigure}
\hfill
\begin{subfigure}{0.48\textwidth}
\centering
\includegraphics[width=\linewidth]{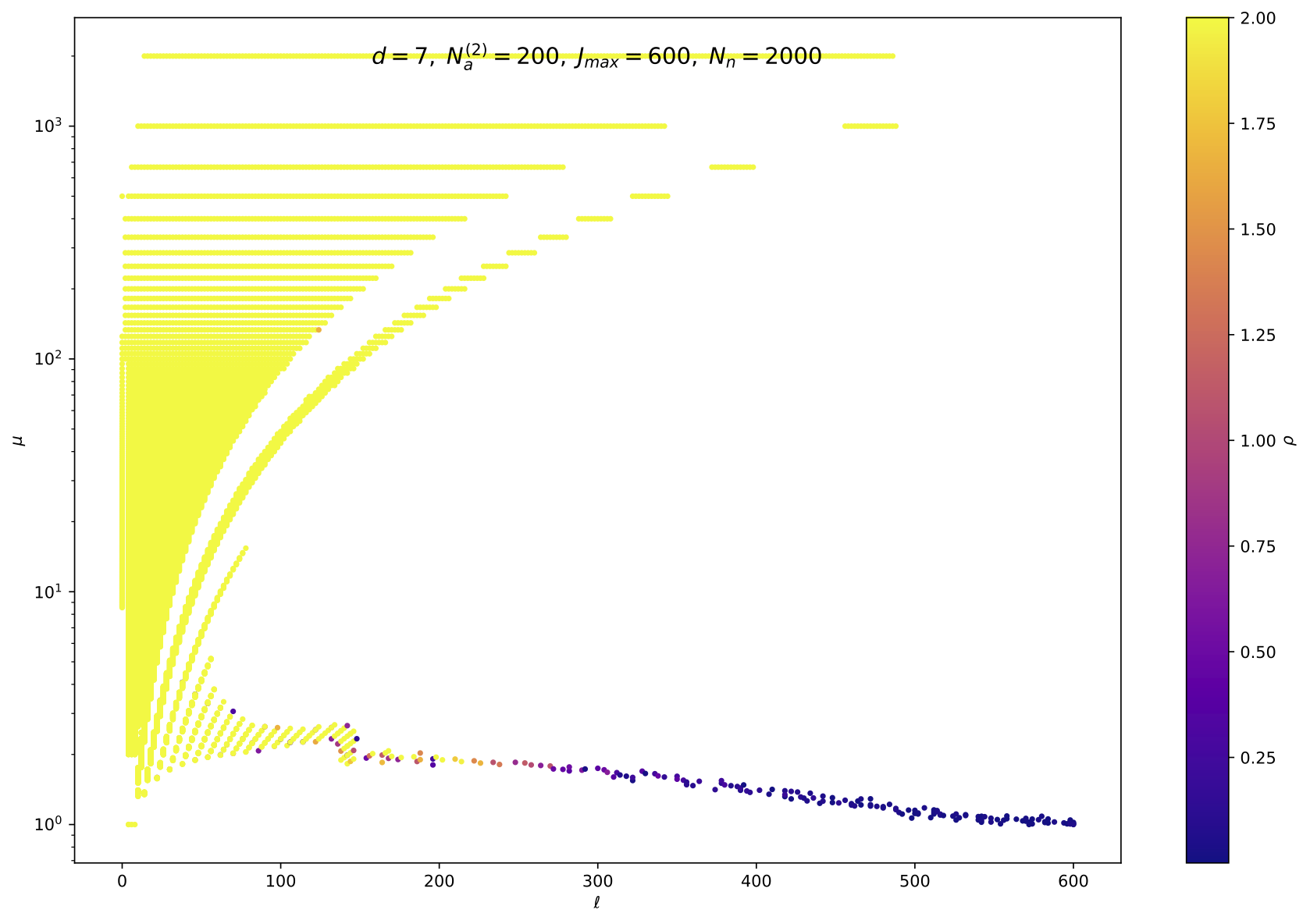}
\caption{$D=7$}
\end{subfigure}
\caption{Extremal spectrum for upper bound on $8\pi G$, using $N_a^{(2)}=200$, $J_{max}=600$, $N_\mu=2000$. The quadratic-band structure persists in higher dimensions, while additional empty regions apeear. In $D>5$ some empty regions are stable with respect to ansatz parameters.}
\label{fig:spectrum_d67}
\end{figure}

Besides the quadratic bands, the extremal spectra also display empty regions, which become even more pronounced in higher dimension, as seen in Fig.~\ref{fig:spectrum_d67}. Unlike the band structure itself, however, these features are more sensitive to the ansatz parameters. As discussed further in Appendix~\ref{app-spectrum}, part of this structure appears to be a truncation artifact, while some empty regions in higher dimensions remain stable.

It is important to emphasize that no string-like assumptions are imposed anywhere in our
analysis for either the projective or non-projective spectra. The emergence of the quadratic structures is therefore a dynamical outcome of simply imposing crossing symmetry, positivity, and linearized unitarity on a gravitational EFT, assuming two-subtraction dispersion relations. In this sense, the result strengthens the connection between dispersive
bootstrap methods and classic expectations from high-energy scattering, while at the same
time pointing to a pattern that is qualitatively different from the linear trajectories more
commonly encountered in string-inspired settings. We leave a detailed and systematic study of these features, as well as a physical interpretation, for future work.

\section{Conclusions and outlook}
\label{sec:discussion}
In this work we developed a primal bootstrap framework for effective field
theories coupled to gravity, based on finite-resolution sampling rather than
smearing. After carefully controlling the numerical instabilities inherent in
the primal formulation, we found that this approach is both robust and
surprisingly informative. On the one hand, it reproduces the known projective
bounds obtained from smearing, thereby providing a strong consistency check on
the method. On the other hand, it naturally gives access to structures that are
much harder to see in the standard dual or smeared formulations, most notably
new non-projective bounds and the extremal spectra that saturate both the
projective and non-projective constraints.

Positivity alone constrains only ratios of Wilson coefficients,
so the allowed region is projective in EFT coupling space. Once
linearized unitarity is imposed, however, this projective structure is
lost. In gravitational EFTs this is especially significant, because the
gravitational coupling itself becomes a physical parameter and absolute
bounds on EFT couplings can arise.

Applying the method to fixed-$a$ dispersion relations in $D=5$, we find a finite upper bound on the dimensionless gravitational coupling,
\[
8\pi G\,M^3 \lesssim O(1)\, ,
\]
equivalently $M/M_{\rm P}\lesssim O(1)$. This indicates that, within the assumptions of
our analysis, the EFT cutoff cannot be taken parametrically above the Planck scale. We
observe the same qualitative non-projective phenomenon in higher dimensions, although the
numerical bounds weaken as the partial-wave normalization grows.

A central advantage of the primal bootstrap is that it gives direct access to
extremal spectra. This is true both for the familiar projective bounds and for
the new non-projective ones. In the projective case, the extremal spectra we
find differ significantly from those inferred in previous dual-bootstrap
analyses, despite leading to very similar bounds. Rather than being concentrated
on a sparse set of isolated states, the spectrum appears to be technically
supported over a broad region of the $(J,\mu)$ plane, but with states heavily
suppressed away from a distinguished lower-right region organized by quadratic
trajectories. In the non-projective case the structure is even sharper: the
extremal spectra organize into narrow quadratic bands, and these bands
themselves follow a further quadratic pattern. In both cases the dominant
organization is strikingly rigid.

These spectra are surprising. They emerge naturally from the extremization
problem, yet they are not of the familiar linear Regge type one might have
expected from a string-like picture. Instead, the dominant structures are well
approximated by quadratic trajectories, and the full extremal solutions exhibit
a nested quadratic organization. Whether this reflects a new class of extremal
gravitational theories, or a universal feature of the bootstrap at finite resolution, remains an open question.

Several important directions remain open. A systematic implementation of
exact unitarity, beyond the linearized approximation, will likely
require semidefinite or nonlinear optimization methods. Incorporating
loop-level gravitational and EFT corrections in a controlled way is
another crucial step, especially in four dimensions where infrared
effects are most subtle. The primal approach is also a natural framework for imposing further constraints, such as the presence or absence of particular states, low-spin dominance \cite{Bern:2021ppb}, or spectral constraints associated with the Completeness Hypothesis \cite{Hillman:2024ouy,Calisto:2025tjo}. It would also be interesting to generalize the
analysis to theories with additional light states, gauge interactions,
or higher-spin fields, and to explore the implications of non-projective
bounds for other swampland conjectures.

\appendix

\section{Numerical implementation and stability tests}
\label{app:numerical_stability}

In this appendix, we present a series of numerical stability tests for the
bounds obtained in the main text and discuss an important limitation
associated with finite spin truncations in the presence of gravity.
\subsection{Illustrative numerical implementation}
\label{app:code_example}

To make our numerical procedure explicit, we present a minimal example of
the linear programming setup used throughout this work. This example
implements the \emph{$k=2$ crossing-symmetric dispersive sum rule only},
imposing \emph{positivity constraints alone} and fixing the overall
normalization. No linearized unitarity constraints or higher-$k$ sum rules
are included, so the resulting bound is projective and serves as a
baseline for comparison.

We discretize the dispersive integral using $N_\mu=300$ energy points and
truncate the partial-wave expansion at $J_{\max}=40$, keeping even spins.
The crossing variable $a$ is sampled at $N_a=20$ Chebyshev points in the
interval $a\in(-1/3,0)$. The spectral density $\rho(z,J)$ is treated as
a set of non-negative variables, and the normalization is fixed by setting
$k_2=1$. The objective is to minimize the EFT coupling $g_2$.

The full optimization problem is a linear program and can be solved
efficiently. An explicit Mathematica implementation is shown below.

\begin{verbatim}
ClearAll[n, P, B, a, eq2];

(* Partial-wave normalization *)
n[l_, d_] := ((4 Pi)^(d/2) (d + 2 l - 3) Gamma[d + l - 3])/
              (Pi Gamma[(d - 2)/2] Gamma[l + 1]);

(* Gegenbauer polynomial *)
P[l_, d_, x_] :=
  Hypergeometric2F1[-l, l + d - 3, (d - 2)/2, (1 - x)/2];

(* k=2 dispersive kernel *)
B[a_, mu_, l_, d_] :=
  (2 mu - 3 a)/mu^3 P[l, d, Sqrt[mu + 3 a]/Sqrt[mu - a]];

(* Discretization parameters *)
NN = 20; nn = 300; jj = 40;

(* Chebyshev sampling in a *)
a[i_] := N[-1/6 + (1/6) Cos[((2 i - 1) Pi)/(2 NN)], 30];

(* Variables *)
varR = Flatten@Table[rho[z, l], {z, 1, nn}, {l, 0, jj, 2}];
vars = Join[{g2, g3, k2}, varR];

(*k=2 sum rule*)
eq2[ai_, d_] := 
  Sum[rho[z, l] B[ai, nn/z, l, d] n[l, 
      d] (1/Pi) (1/nn) (z/nn)^(d/2 - 3), {l, 0, jj, 2}, {z, 1, nn}] + 
   k2/ai - 2 g2 + g3 ai;

(* Constraints: sum rules + positivity + normalization *)
cons[d_] :=
  Join[(eq2[#, d] == 0) & /@ Table[a[i], {i, 1, NN}],
       Thread[varR >= 0],
       {k2 == 1}];

(* Solve LP *)
g2 /. LinearOptimization[g2, cons[6], vars,
      Method -> "CLP", Tolerance -> 10^-12]
\end{verbatim}

For $D=6$, this program yields the bound
\begin{equation}
\frac{g_2}{8 \pi G}>-15.6864 ,
\end{equation}
in agreement with the projective bounds reported in the main text.

We remark that a Mathematica implementation for higher ansatz parameters becomes unstable, in particular for the high spin cutoff needed for fixed-$t$ analysis.
The complete numerical code and more efficient Python implementation used for most results in this paper can be made available by the authors upon request.

\subsection{Stability and convergence tests: fixed-$t$ \label{fix_t_test}}

In this subsection, we perform numerical convergence tests for projective bounds using the fixed-$t$ dispersion relation.

\begin{figure}[ht]
\centering

\begin{subfigure}{0.48\textwidth}
    \centering
    \includegraphics[height=5cm]{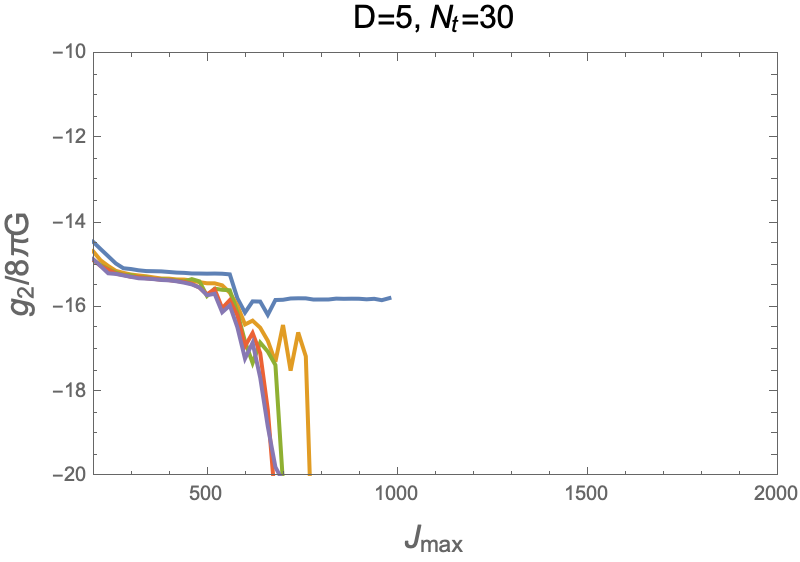}
\end{subfigure}
\hfill
\begin{subfigure}{0.48\textwidth}
    \centering
    \includegraphics[height=5cm]{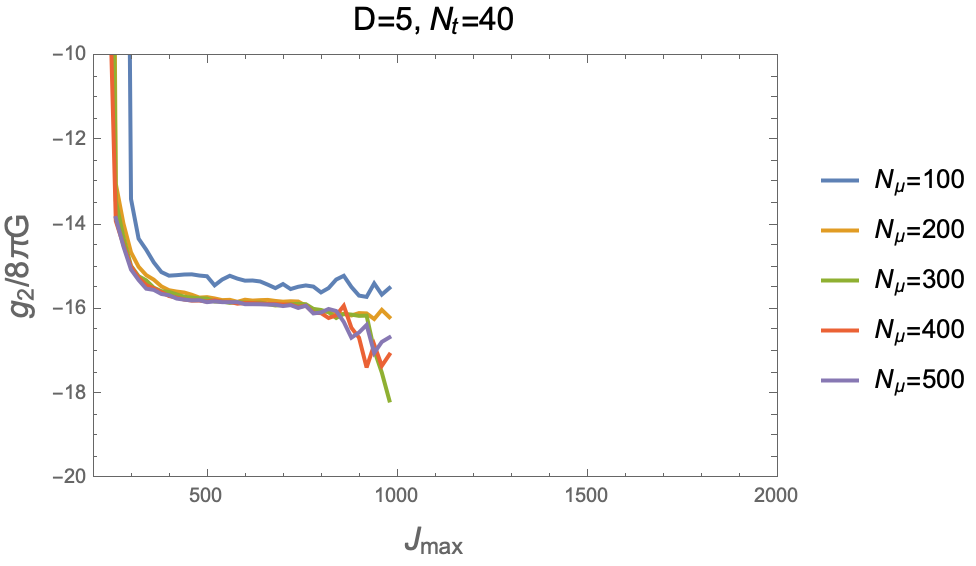}
\end{subfigure}

\vspace{0.5cm}

\begin{subfigure}{0.48\textwidth}
    \centering
    \includegraphics[height=5cm]{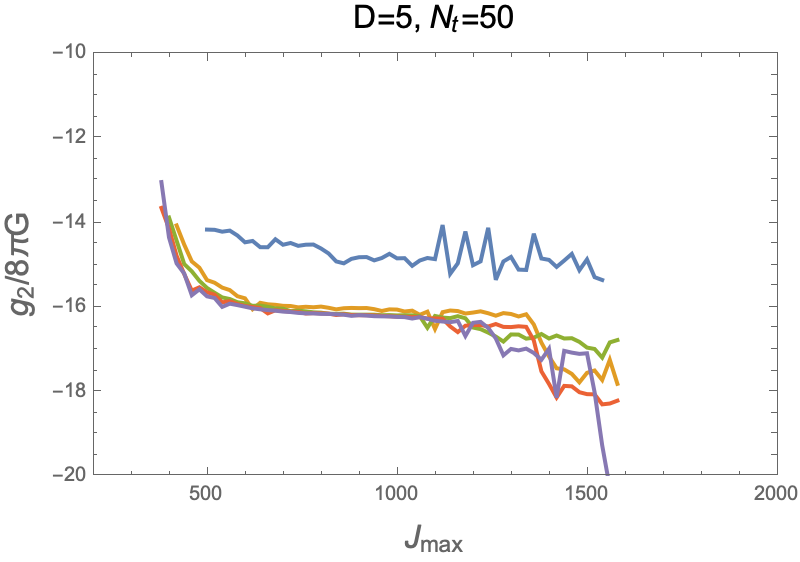}
\end{subfigure}
\hfill
\begin{subfigure}{0.48\textwidth}
    \centering
    \includegraphics[height=5cm]{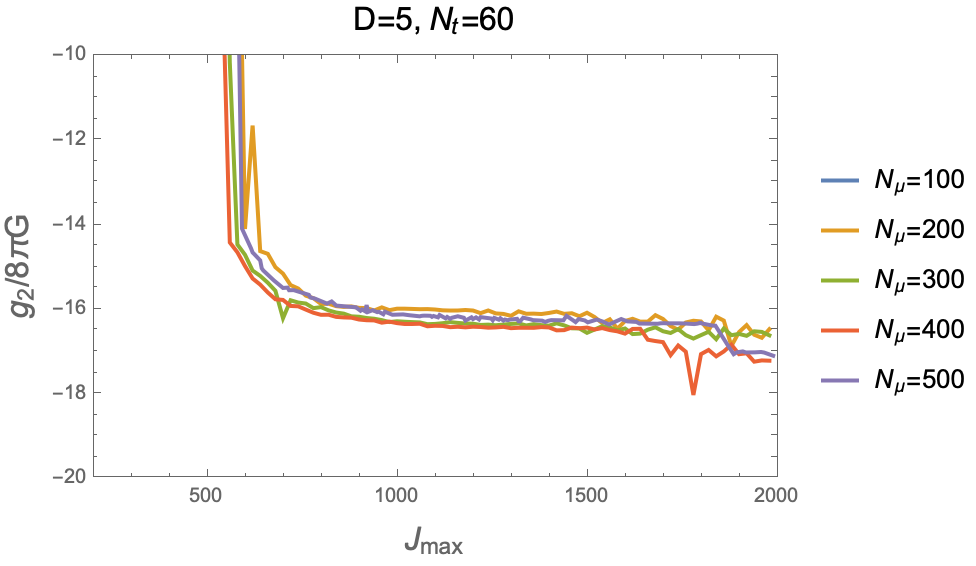}
\end{subfigure}

\caption{Comparing convergence in $N_\mu$ and $N_t$ in $D=5$. Convergence in energy discretization is reached around $N_\mu=500$. Increasing $N_t$ leads to wider stable window of bounds}
\label{fig:d=5testA}
\end{figure}

\begin{figure}[ht]
\centering
    \includegraphics[height=7cm]{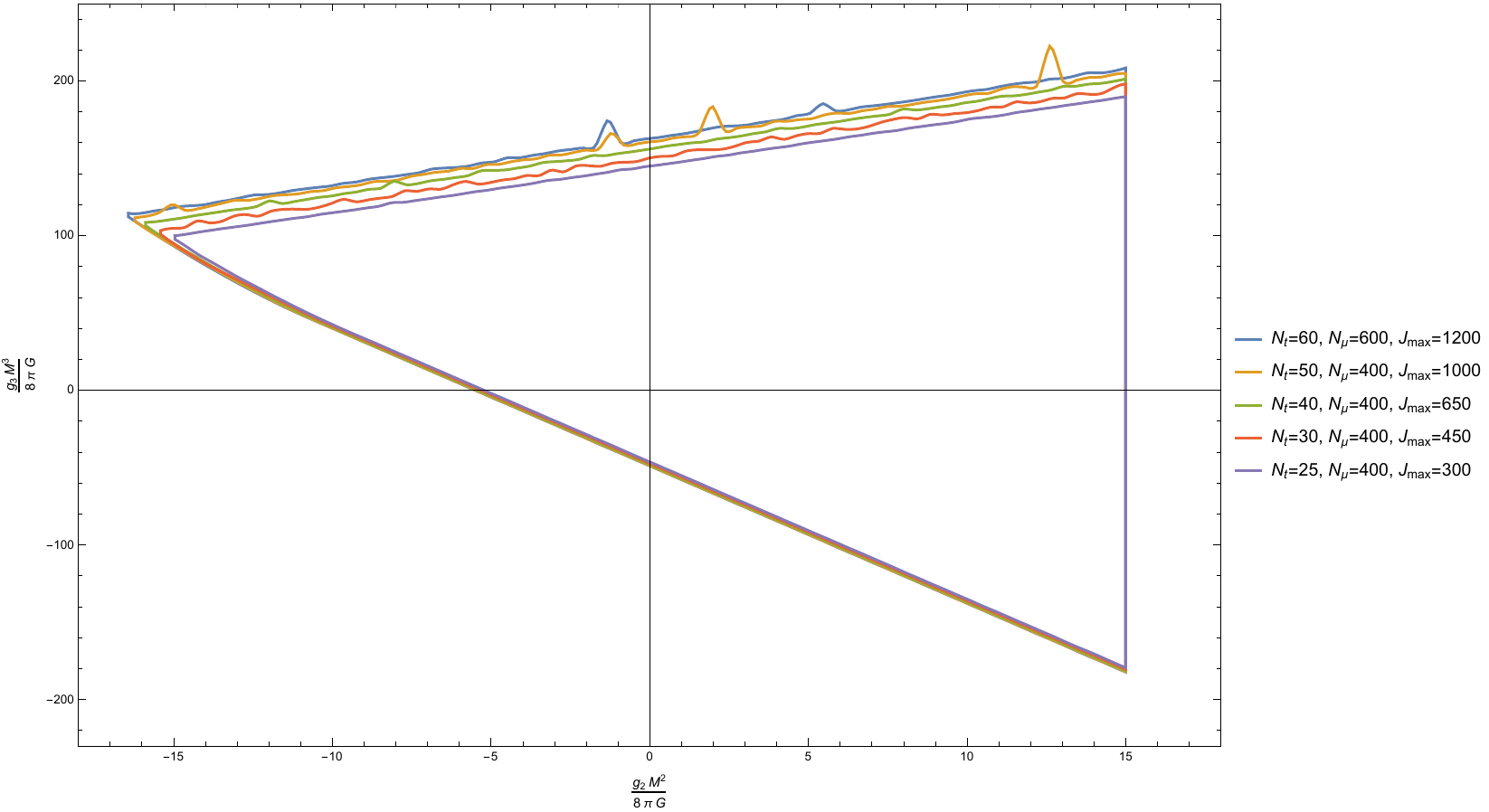}
    \caption{Convergence for bounds in the 2D plot ($g_2/8\pi G,g_3/8\pi G$) in $D=5$. The upper boundary is slower to converge, and exhibits minor numerical instabilities.}
\label{fig:g32plot}
\end{figure}

\begin{figure}[ht]
\centering
    \includegraphics[height=6cm]{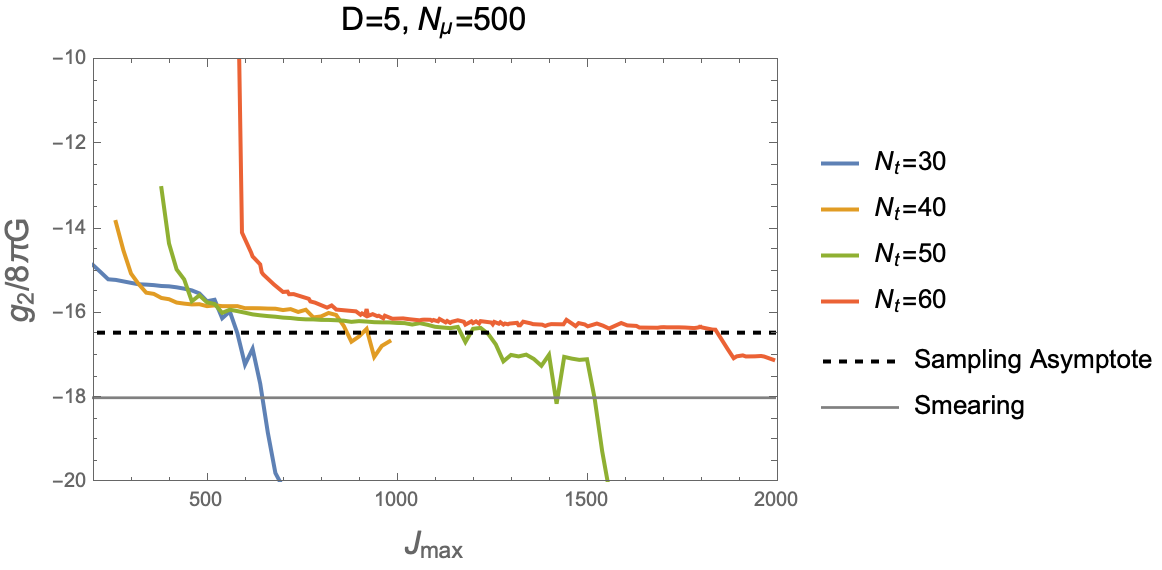}
      \caption{Convergence in $N_t$ at high $N_\mu$ asymptotes to a fixed value, leading to a stronger bound compared to smearing in $D=5$.}
\label{fig:d=5testB}
\end{figure}

\begin{figure}[ht]
\centering

\begin{subfigure}{0.48\textwidth}
    \centering
    \includegraphics[height=5cm]{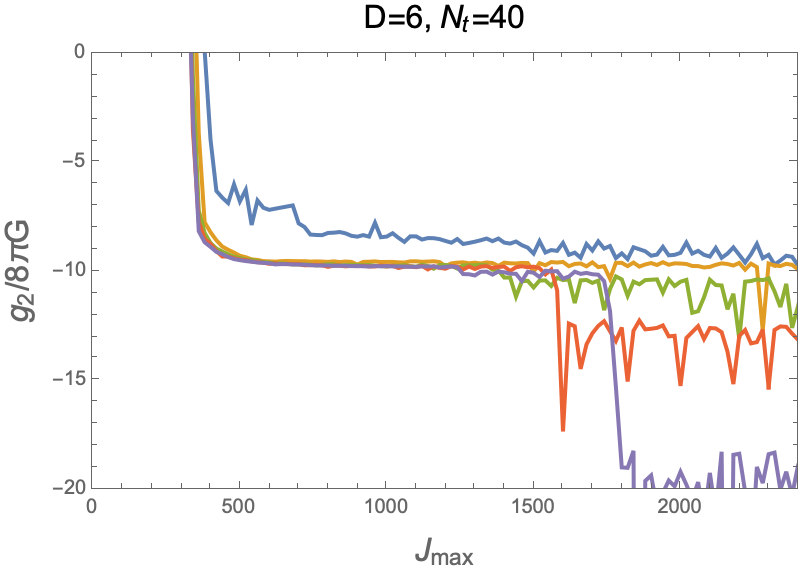}
\end{subfigure}
\hfill
\begin{subfigure}{0.48\textwidth}
    \centering
    \includegraphics[height=5cm]{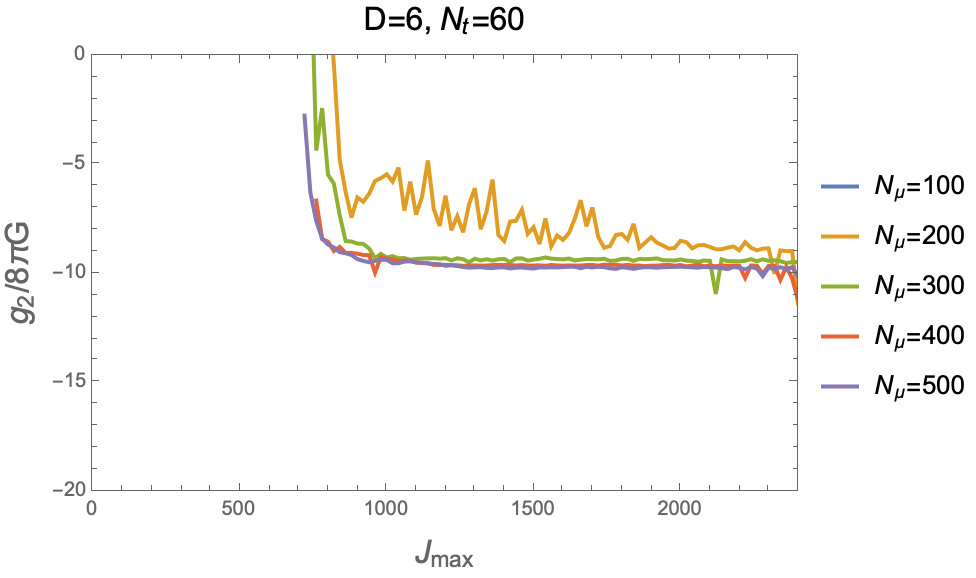}
\end{subfigure}
\caption{Comparing convergence in $N_\mu$ and $N_t$ in $D=6$. At higher $N_t$ the stable window increases, and numerical instabilities are reduced by also increasing energy discretization $N_\mu$.}
\label{fig:d=6testA}
\end{figure}

\begin{figure}[ht]
\centering
    \includegraphics[height=6cm]{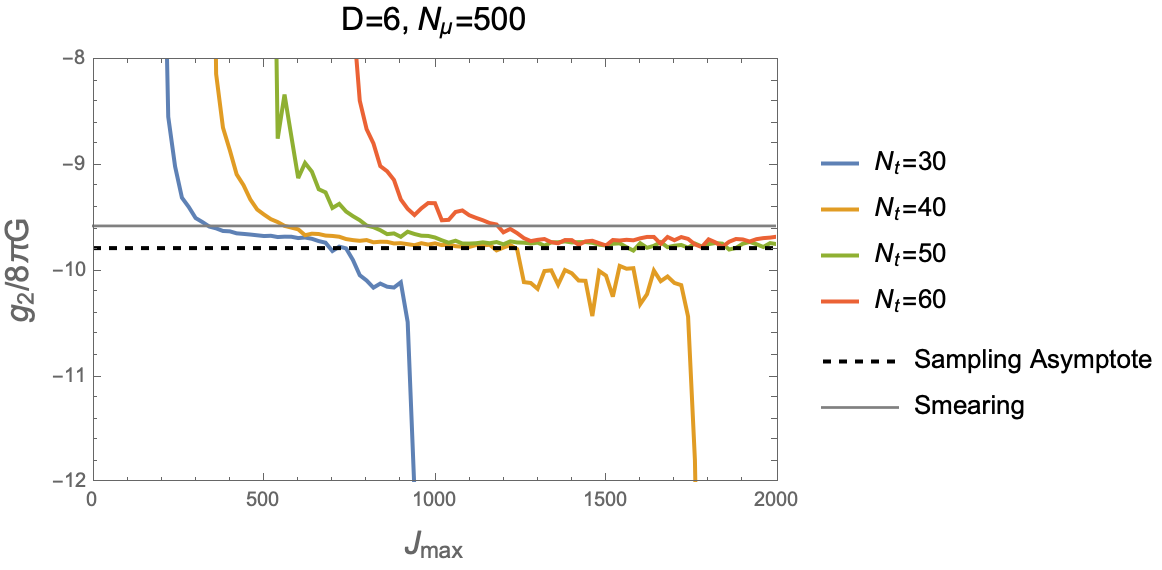}
    \caption{Convergence in $N_t$ at high $N_\mu$ asymptotes to a fixed value, leading to a slightly weaker bound compared to smearing in $D=6$.}
\label{fig:d=6testB}
\end{figure}

From Figs.~\ref{fig:d=5testA} and \ref{fig:d=5testB} (for $D=5$) and Figs.~\ref{fig:d=6testA} and \ref{fig:d=6testB} (for $D=6$) for the lower bound on $g_2/8\pi G$; and Fig.~\ref{fig:g32plot} for the 2D plot ($g_2/8\pi G,g_3/8\pi G$) we observe the following
\begin{itemize}
\item At fixed $N_t$, there exists a window defined by $r_\textrm{min}(t)N_t<J_{\textrm{max}}<r_\textrm{max}(t)N_t$ within which the bounds are numerically stable with respect to $J_{\textrm{max}}$. As $N_t$ is increased, this window increases, and becomes more convergent. Unlike the fixed-$a$ case to be studied below, the ratios $r$ increase with $N_t$. For example at $N_t=40$ the stable window is $10 N_t\lesssim J_{\textrm{max}}\lesssim 18 N_t$, while for $N_t=60$ it is $16 N_t\lesssim J_{\textrm{max}}\lesssim 30 N_t$
\item The bounds mostly converged around $N_t\lesssim 60$, $N_\mu\lesssim 500$, $J_\textrm{max}\lesssim 2000$
\item For particular combinations of ansatz parameters, some local instabilities persist, likely due to accidental degeneracies in the numerics
\end{itemize}

\subsection{Stability and convergence tests: fixed-$a$ \label{A.3}}
In this section we investigate the numerical stability of our approach using fixed-$a$ dispersion relations. 

\begin{figure}[htbp] 
   \centering
   \includegraphics[width=4in]{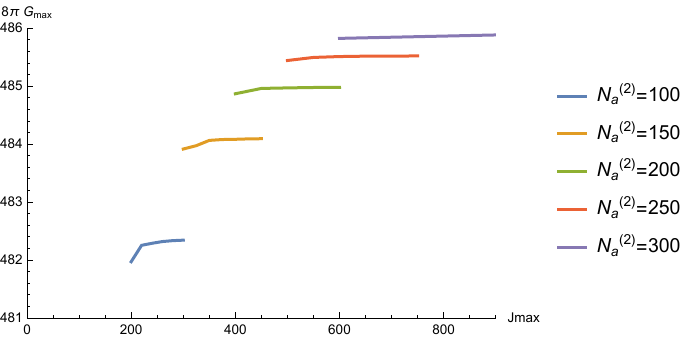} 
   \caption{Stable bounds are obtained for $2 N_a^{(2)} \lesssim J_{\textrm{max}} \lesssim 3 N_a^{(2)}$. Below the minimal spin a solution does not exist, above the maximal spin the bound diverges. Within this region bounds rapidly converge with increasing $N_a^{(2)}$ }
   \label{fig:stable_region}
\end{figure}

\begin{figure}[htbp] 
   \centering
   \includegraphics[width=4in]{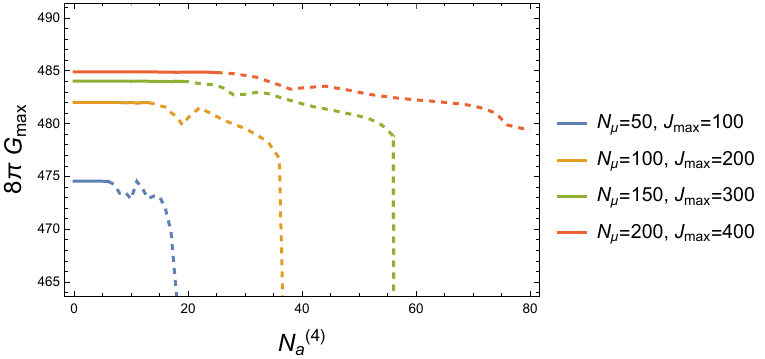} 
   \caption{At fixed ratio $J_{\textrm{max}}/N_a^{(2)}$, the bounds are stable and virtually  unaffected (the change is less than $0.01\%$) if we keep  $N_a^{(4)}\lesssim N_a^{(2)}/8$, shown as the solid lines in the figure. When $N_a^{(4)}\gtrsim N_a^{(2)}/3$, the bound quickly drops off, meaning the ansatz space is too small.}
   \label{fig:Na4}
\end{figure}

\begin{figure}[htbp] 
   \centering
   \includegraphics[width=4in]{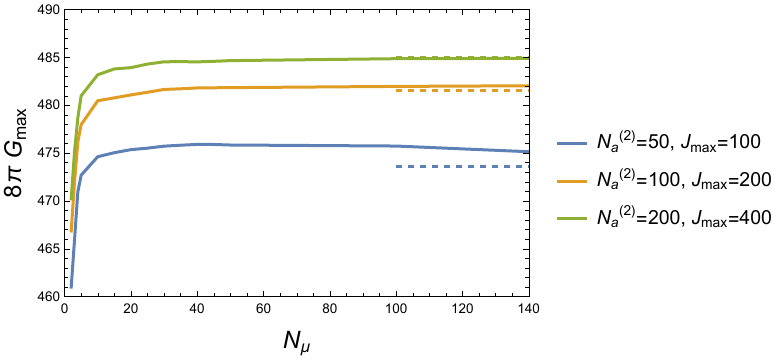} 
   \caption{At fixed ratio $J_{\textrm{max}}/N_a^{(2)}$, the bounds converge rapidly with discretization in $N_\mu$. The convergence is even faster with increasing $N_a^{(2)}$. We plot $J_{\textrm{max}}/N_a^{(2)}=2$, but the result is similar for any ratio within the stable region $2\lesssim J_{\textrm{max}}/N_a^{(2)}\lesssim 3$. Even at $N_\mu=2$, the bound is within $3\%$ of the asymptotic value at $N_\mu\rightarrow \infty$.}
   \label{fig:stable_Nn}
\end{figure}

We find the following main features:
\begin{itemize}
\item Stable bounds require $2 N_a^{(2)} \lesssim J_{\textrm{max}} \lesssim 3 N_a^{(2)}$, and converge rapidly with increasing $N_a^{(2)}$ within this region. Figure \ref{fig:stable_region}
\item Higher order dispersion relations have negligible effect on the bounds, but need to be truncated, for instance we require $N_a^{(4)}\lesssim J_{max}/16$. For higher $N_a^{(4)}$, the bound becomes unstable. Figure \ref{fig:Na4}
\item Bounds converge rapidly with increasing $N_\mu$. Figure \ref{fig:stable_Nn}
\end{itemize}

 The stability region  $2 N_a^{(2)} \lesssim J_{\textrm{max}} \lesssim 3 N_a^{(2)}$ (or the one observed for fixed-$t$) can be understood as follows. The lower bound reflects the fact that, for a fixed number of parameters $N_a$, no solution exists unless sufficiently high spin states are included. A similar phenomenon already appears in the case without the graviton pole, where it is known that null constraints require contributions from increasingly high spin states \cite{Chiang:2021ziz}. In the present case this effect is significantly amplified by the presence of the graviton pole. Indeed, it is known that the graviton pole itself requires an infinite tower of higher-spin states \cite{Haring:2024wyz}. Consequently, if $J_{\textrm{max}}$ is chosen too small relative to $N_a$, the constraints cannot be satisfied.

On the other hand, if the maximal spin is taken too large compared to $N_a$, the numerical bounds begin to diverge. This behavior is not physical but rather an artifact of the sampling approach used here. In contrast, forward-limit methods based on the primal bootstrap are known to converge as the maximal spin increases. Ordinarily such lack of convergence would indicate a limitation of the numerical method. However, in our case we observe a clear stability window in which the bounds remain completely flat as $J_{\textrm{max}}$ is increased. Importantly, this stability region grows with both $N_a$ and $J_{\textrm{max}}$. The existence of this plateau means that the resulting bounds are reliable. As $N_a \to \infty$ and $J_{\textrm{max}} \to \infty$, the precise value of $J_{\textrm{max}}$ becomes irrelevant, and the bounds become independent of the details of the ansatz. In this limit (which in practice is reached quickly) the results are therefore robust.

Taken together, these tests indicate that the upper bound on
$(8\pi G)$ is stable under simultaneous refinements of the energy,
spin, and kinematic discretizations, provided that the sampling in the
kinematic variable $a$ is increased in a manner consistent with the spin truncation.

\subsection{Extremal spectrum stability}\label{app-spectrum}
\begin{figure}[t]
\centering
\includegraphics[width=0.8\textwidth]{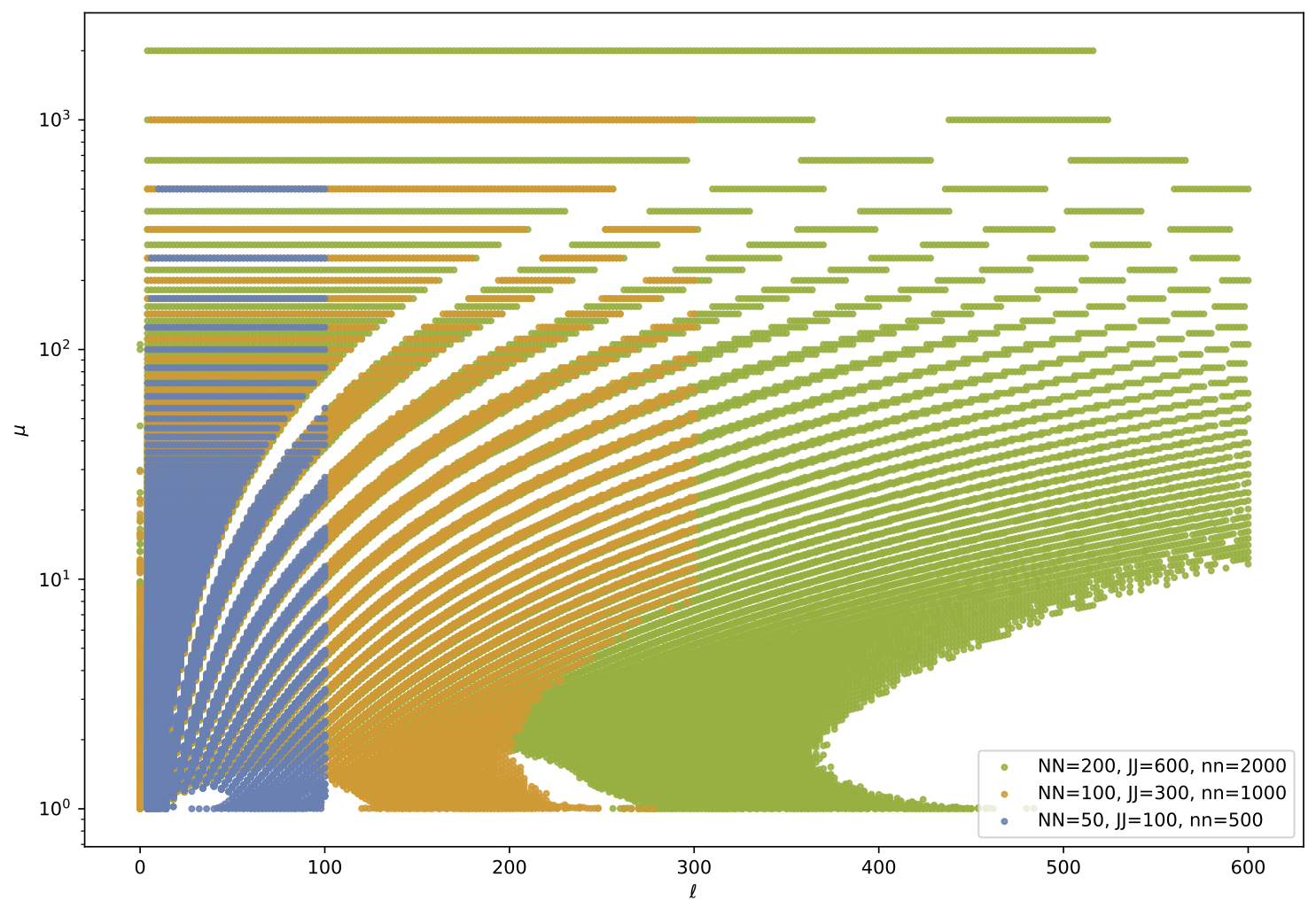}
\caption{
The shape of the quadratic trajectories is stable with refining ansatz parameters. However, the empty regions extend proportionally.
}
\label{fig:spectrum_convergence}
\end{figure}

In this subsection we present the extremal spectrum plots corresponding to the upper bound on $8\pi G$. In Fig.~\ref{fig:spectrum_convergence}, we show the spectra for different ansatz parameters. 

\begin{figure}[t]
\centering
\includegraphics[width=0.8\textwidth]{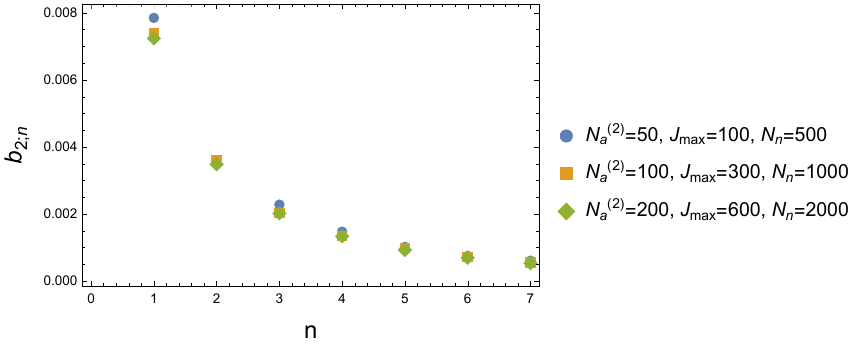}
\caption{
The shape of the trajectories converges rapidly with refining ansatz parameters. Shown are the $b_{2;n}$ coefficients of the $J^2$ terms in the first few leading trajectories.
}
\label{fig:b2_convergence}
\end{figure}

The trajectory component of the spectrum is stable under increasing ansatz parameters, with the leading coefficients $b_{2;n}$ rapidly converging, as shown in Fig.~\ref{fig:b2_convergence}. In contrast, the empty regions appear to be truncation artifacts, as their size grows proportionally with the maximum spin cutoff.

\section{Loop level corrections}
\label{loopchecks}

In this subsection, we estimate the loop-level corrections to the amplitude for the extremal value of $8\pi G$. As discussed in the main text, including nonlinear contributions such as $(8\pi G)^2$, $g_2^2$, and $g_2 g_3$ renders the optimization problem nonlinear, preventing a direct determination of the boundary of an individual coupling.

\subsection{Graviton loops.}
To simplify the analysis, we temporarily neglect the EFT loop contributions and focus solely on the gravitational loop effects, in order to isolate their impact on the bounds. In this setup, once the value of $k_2=8\pi G$ is fixed, the problem reduces again to a linear optimization problem. A full treatment including EFT loop contributions is left for future work.

We follow the amplitude formulation of Ref.~\cite{Chang:2025cxc} and, for simplicity, retain only the contribution proportional to $(\kappa^2)^2$.

\begin{equation}
   \mathcal{M}_{\text{one-loop}}= \mathcal{M}_{\text{one-loop}}^{\text{grav-grav}}+\cdots
\end{equation}

\begin{equation}
\begin{aligned}
c_2^{\text{grav-grav}}\big|_{d=6}
&=
\frac{(\kappa^2)^2}{107520\,a\,\pi^4}
\Bigg[
2126 a M^2 + 700 a^2 \pi^2
+ 630 a^2 \log^2\!\left(-\frac{a}{M^2}\right)
\\
&\quad
- 1409 a^2 \log\!\left(1-\frac{M^2}{a}\right)
+ 420 M^4 \log\!\left(1-\frac{M^2}{a}\right)
- 525 a^2 \log^2\!\left(1-\frac{M^2}{a}\right)
\\
&\quad
- 420 M^4 \sqrt{\frac{3a+M^2}{M^2-a}}
\log\!\left(
\frac{1+\sqrt{\frac{3a+M^2}{M^2-a}}}
{1-\sqrt{\frac{3a+M^2}{M^2-a}}}
\right)
\\
&\quad
+ 945 a^2
\log^2\!\left(
\frac{1+\sqrt{\frac{3a+M^2}{M^2-a}}}
{1-\sqrt{\frac{3a+M^2}{M^2-a}}}
\right)
+ 1260 a^2 \operatorname{Li}_2\!\left(\frac{a}{M^2}\right)
\Bigg],
\end{aligned}
\end{equation}
where we define $\kappa^2 = 8\pi G$.

\vspace{0.5em}

To proceed, we fix $k_2$ and determine the corresponding $g_2$ boundary in $D=6$, enabling a direct comparison with tree-level results. Fig.~\ref{fig:k2g2loop6d} illustrates how the allowed region deforms at both tree and loop level depending on the parameters.
\begin{center}
\includegraphics[width=0.80\textwidth]{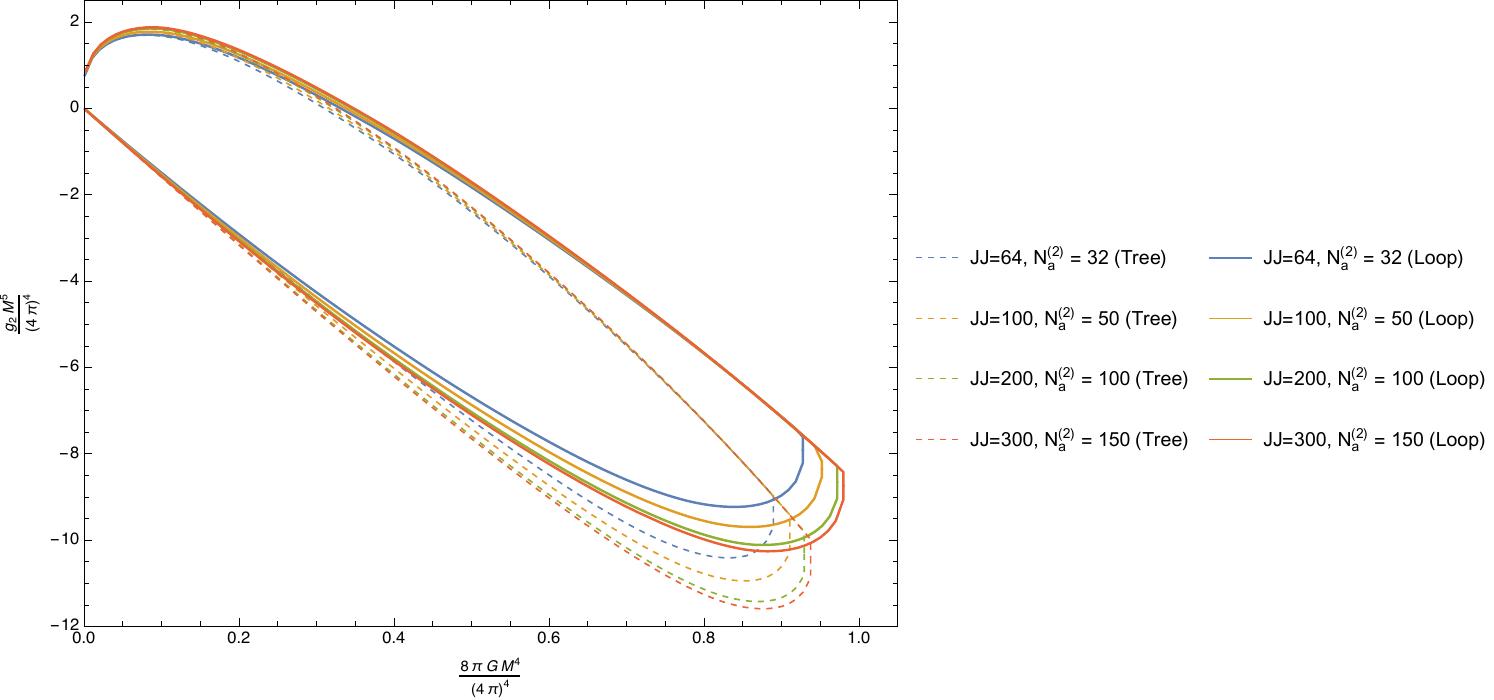}
\captionof{figure}{
Allowed regions in the 
$\bigl( (8\pi G)/(4\pi)^{4},\, g_2/(4\pi)^{4} \bigr)$ plane 
for gravity-loop contributions in flat spacetime in $D=6$ with heavy mass scale $M$,
comparing tree-level and one-loop results under linear unitarity conditions. 
}
\label{fig:k2g2loop6d}
\end{center}

We find that the tree-level and one-loop results agree well in the weak-coupling regime. As $8\pi G$ increases, the bound changes only mildly, indicating that it is not significantly affected by gravitational loop corrections.

\vspace{0.5em}

Another regime of interest is obtained by comparing with Fig.~\ref{fig:g3g2_gravity} after including loop-level amplitudes.

\begin{center}
\includegraphics[width=0.80\textwidth]{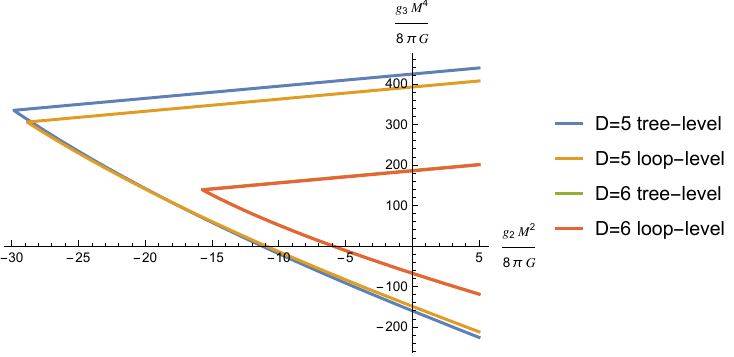}
\captionof{figure}{
Allowed regions in the 
$\bigl( g_2/(8\pi G),\, g_3/(8\pi G) \bigr)$ plane in $D=5$ and $D=6$, comparing tree-level and one-loop results at fixed value $8\pi G=1$. 
}
\label{fig:g2g3loop}
\end{center}

We find that, after including loop-level amplitudes, the allowed region in the $(g_2,g_3)$ plane changes only slightly in $D=5$, and remains nearly unchanged in $D=6$, largely due to the larger normalization factor. This suggests that the tree-level bounds on $(g_2,g_3)$ in the weak-coupling region remain robust even after incorporating loop-level corrections.

\subsection{EFT loops}

\begin{center}
\label{loop}
\includegraphics[width=0.7\textwidth]{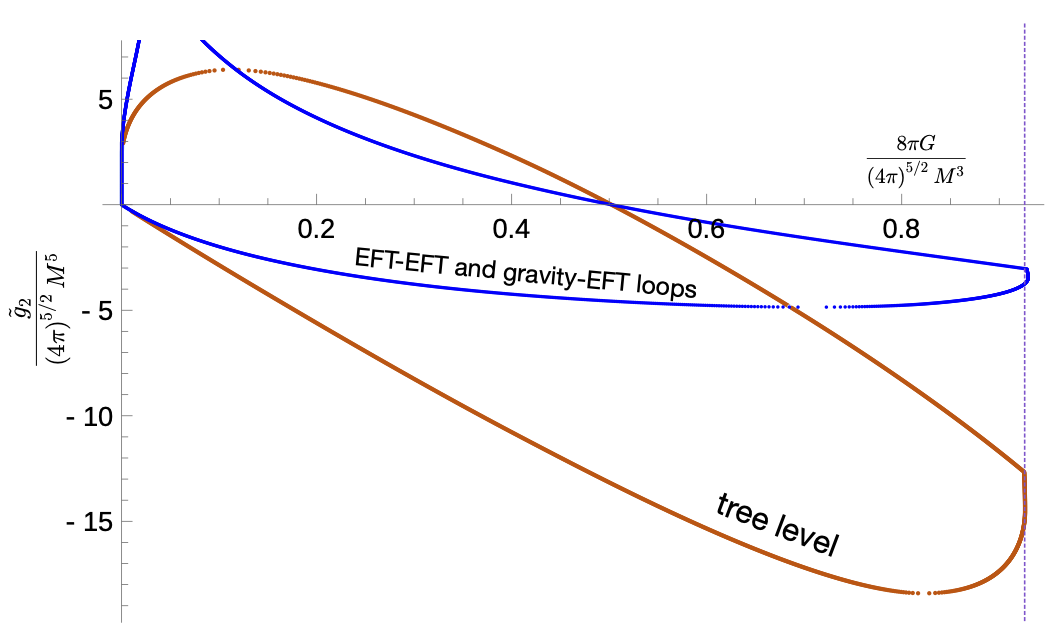}
\captionof{figure}{
Allowed regions for $g_2$ and $8\pi G$ for the loop corrections \eqref{EFTloop}  in the case of $N=1$. The overall upper bound on $8\pi G$ is almost not affected, although the region shape for $g_2$ is significantly deformed.
}
\label{fig:g2g3loop}
\end{center}

EFT loop corrections can in principle also modify the non-projective bounds on $g_2$ and
$8\pi G$, since the crossing-symmetric dispersion relations receive
non-analytic contributions from EFT loops.
To illustrate this effect,
we restrict to the leading loop terms generated by $g_2$. In large-$N$ $O(N)$
models, one expects EFT loops involving $g_3$ and higher couplings to be
parametrically suppressed, so this provides a natural first approximation.
Accordingly, in this subsection we include only the leading $g_2$-dependent EFT
loops and study how they modify the numerical bounds.

We consider the scattering of identical scalars in a five-dimensional $O(N)$
model with an additional shift symmetry, described at tree level by the
crossing-symmetric amplitude
\begin{equation}
A(s,t)=\frac{8\pi G \,(s^2+t^2+u^2)^2}{stu}
+\frac{\tilde g_2}{M^5}(s^2+t^2+u^2)
+\frac{\tilde g_3}{M^7}stu+\dots.
\end{equation}
The leading EFT-EFT and EFT-gravity loop corrections generated by $\tilde g_2$
can be computed from unitarity cuts \cite{inpreparation,Guerrieri:2020bto}. In
five dimensions they take the form
\begin{equation}
\label{EFTloop}
\begin{split}
A_{\text{loop}}=&
-\frac{\tilde g_2 (8\pi G)\sqrt{-s}\,s\left((69N+4173)s^2-16(N-128)st-16(N-128)t^2\right)}
{49152\pi M^5}
\\[4pt]
&+\frac{\tilde g_2^2 \sqrt{-s}\,s^2\left(3(32N+53)s^2+8st+8t^2\right)}
{98304\pi M^{10}}
+\dots,
\end{split}
\end{equation}
where the omitted terms include loop contributions involving $\tilde g_3$ and
higher EFT couplings.

We incorporate these corrections into the crossing-symmetric dispersion relation
by modifying the low-energy side of the $k=2$ sum rule,
\begin{equation}
-\frac{8\pi G}{a}+2\tilde g_2-a\tilde g_3
-\frac{1}{2\pi i}\oint_{\Gamma_{\rm low}}
A_{\text{loop}}(\mu,\tau(\mu,a))\,
\frac{2\mu-3a}{\mu^4}\,d\mu
=
\int_{M^2}^{\infty}(\dots)\,.
\end{equation}

The resulting bounds are shown in Fig.~\ref{fig:g2g3loop}. Remarkably, although
the inclusion of the $g_2$ loop corrections significantly deforms the allowed
region in $g_2$, the upper bound on $8\pi G$ remains almost unchanged. This is
somewhat surprising, and suggests that the bound on $8\pi G$ may be considerably
more robust than the detailed shape of the allowed EFT parameter space. A full
analysis including all EFT loop effects is left for future work.

\section{Results in four dimensions with gravity}
\label{sec:4d_gravity}

We now turn to the four-dimensional case with gravity. In contrast to
smearing-based approaches, where one faces a tension between maintaining
positivity at large impact parameter while simultaneously regulating the
gravitational pole, our functional implementation does not encounter an
immediate obstruction at the level of the crossing-symmetric dispersion
relation. A closely related effect does, however, reappear through the
finite-spin truncation used in the numerics, as we discuss below. 


\label{subsec:4d_projective}

In the smearing framework, requiring positivity at large impact
parameter $b$ while regulating the gravitational pole leads to a tension.
A practical resolution is to introduce an explicit upper bound on $b$,
which effectively imposes an infrared cutoff by restricting the range of
the Fourier-conjugate variable. Since large impact parameter corresponds
to small momentum transfer, this procedure regulates the infrared
divergence associated with graviton exchange. As a result, the lower
bound on $g_2/(8\pi G)$ acquires a characteristic logarithmic dependence
on the impact-parameter cutoff.

In our setup, no explicit regulator is introduced at the level of the
crossing-symmetric sum rules. However, an analogous effect arises once
the partial-wave expansion is truncated at finite spin. Using the
relation
\begin{equation}
b \sim \frac{2J}{\sqrt{s}},
\end{equation}
restricting to $J \le J_{\max}$ effectively imposes an upper bound
$b_{\max} \sim 2J_{\max}$. Therefore, although the technical
implementations differ, both approaches encode the same physical
limitation: resolving long-range gravitational interactions requires
access to arbitrarily large impact parameters, or equivalently,
arbitrarily high angular momenta.

To quantify this effect, we study the dependence of the projective bound
on $g_2/(8\pi G)$ as a function of $J_{\max}$. We fix the energy
discretization to $N_\mu=300$ and scale the number of sampling points as
\begin{equation}
N_a^{(2)}=\frac{J_{\max}}{2},
\qquad
N_a^{(4)}=\frac{J_{\max}}{8}.
\end{equation}
The resulting bounds are presented as a function of $\log b_{\max}$.

\paragraph{Lower bound on $g_2/(8\pi G)$ and its dependence on $J_{\max}$.}

\begin{table}[t]
\centering
\begin{tabular}{c|c|c}
\hline
$J_{\max}$ & $\log b_{\max}$ & $(g_2/8\pi G)_{\min}$ \\
\hline
24 & $\log 48$  & $-59.80$ \\
32 & $\log 64$  & $-69.70$ \\
40 & $\log 80$  & $-77.46$ \\
48 & $\log 96$  & $-83.76$ \\
56 & $\log 112$ & $-88.81$ \\
64 & $\log 128$ & $-93.29$ \\
72 & $\log 144$ & $-97.33$ \\
\hline
\end{tabular}
\caption{
Lower bound on $g_2/(8\pi G)$ in four dimensions as a function of
$J_{\max}$ (equivalently $\log b_{\max}$ with $b_{\max}\sim2J_{\max}$).
}
\label{tab:projective_g2_over_G_4d}
\end{table}

\begin{figure}[t]
\centering
\includegraphics[width=0.75\textwidth]{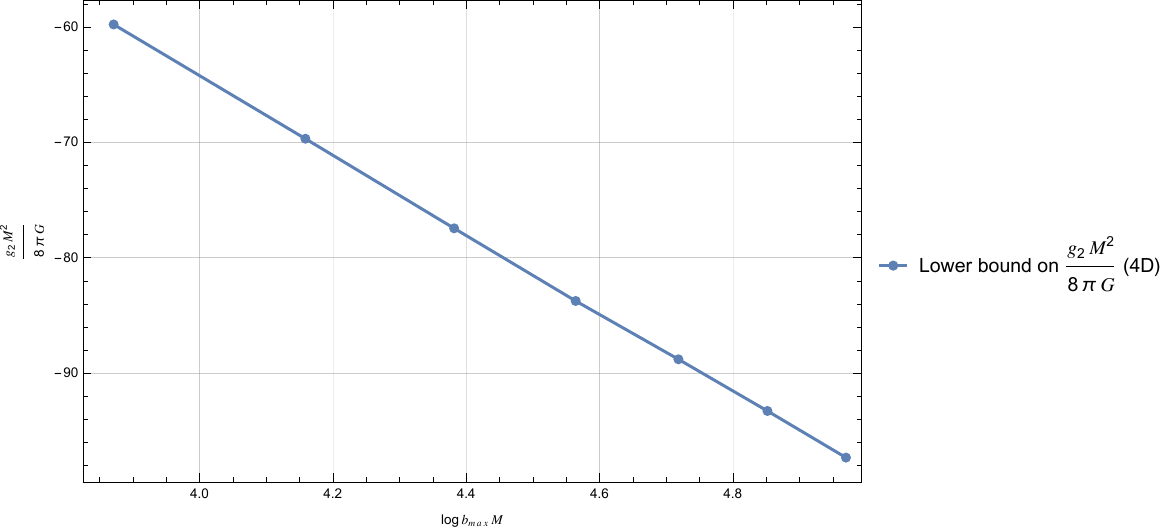}
\caption{
Lower bound on $g_2/(8\pi G)$ in four dimensions as a function of
$\log b_{\max}$. The approximately linear dependence indicates a
logarithmic sensitivity to the large-impact-parameter regime.
}
\label{fig:g2lower4d_logb}
\end{figure}

A linear fit yields
\begin{equation}
g_2 \;\gtrsim\; 8\pi G\,M^2
\left(72.24 - 34.1351\,\log(b_{\max}M)\right),
\end{equation}
over the range of spin cutoffs considered. This bound is weaker than the
fixed-$t$ result obtained using smearing techniques
\cite{Caron-Huot:2021rmr}, but is numerically close to the bound derived
from the fixed-$a$ dispersion relation \cite{Chang:2025cxc}.

\vspace{0.5em}

The above behavior admits a simple physical interpretation. In theories
with massless graviton exchange, dispersive kernels receive dominant
contributions from arbitrarily large impact parameters. Any finite spin
truncation therefore leaves a residual dependence on $b_{\max}$, which
manifests as the observed logarithmic scaling. From this perspective,
the logarithmic behavior is not a numerical artifact, but a direct
reflection of unresolved large-$b$ contributions associated with the
gravitational $t$-channel pole.

This also clarifies the relation to smearing-based approaches. Smearing
implements an explicit infrared regulator in impact-parameter space,
while the present functional method introduces an implicit cutoff
through finite $J_{\max}$. Despite these technical differences, both
approaches capture the same infrared physics, and reliable bounds in the
presence of gravity require control over the large-impact-parameter
regime.

\acknowledgments

YX thanks Celina Pasiecznik for insightful discussions at \emph{Amplitudes 2025}, which motivated the use of fully
crossing-symmetric dispersion relations in this work.  YX acknowledges the online lectures at \emph{TASI 2025} and subsequent discussions with Simon Caron-Huot during the Winter School on \emph{New Frontiers of Quantum Field Theory and Gravity} in Beijing. YX and GZP are very grateful to the committee of Bootstrap 2025 in Kyoto for support during the conference where part of this work was done. The authors thank Denis Karateev for valuable discussions on numerical bootstrap methods and optimization, and Long-Qi Shao for a careful reading of the draft and feedback. LR is supported by the National Natural Science Foundation of China (NSFC) General Program No.~12475070 and by the Beijing Natural Science Foundation International Scientist
Project No.~IS24014. The work of AT was supported by the National Natural Science Foundation of China (NSFC) under Grant No. 12547104 and No. 12505091. 





\bibliographystyle{JHEP}
\bibliography{biblio}
\end{document}